\documentclass[%
readprint,
twocolumn,
superscriptaddress,
showpacs,preprintnumbers,
nofootinbib,
aps,
prd
]{revtex4-1}

\usepackage{subfigure}

\usepackage{amsmath}
\usepackage{amssymb}
\usepackage{mathtools}
\usepackage{hyperref}
\usepackage{placeins}
\usepackage{graphicx}

\usepackage{color}

\usepackage{ulem}

\usepackage{comment}

\def\be{\beta} 
\def\ga{\gamma}

\def\ta{\tau}

\def\De{\Delta}

\def\Om{\Omega}

\newcommand{\ben}{\begin{equation}}
\newcommand{\een}{\end{equation}}
\newcommand{\bea}{\begin{eqnarray}}
\newcommand{\eea}{\end{eqnarray}}
\newcommand{\ba}{\begin{array}}
\newcommand{\ea}{\end{array}}
\newcommand{\bit}{\begin{itemize}}
\newcommand{\eit}{\end{itemize}}

\newcommand{\half}{\frac12}

\newcommand{\bx}{\textbf{x}}

\newcommand{\Hc}{H_*} 
\newcommand{\Nb}{N_\text{b}} 
\newcommand{\Rc}{R_\text{c}} 
\newcommand{\Rstar }{R_*} 
\newcommand{\vw}{v_\text{w}} 

\newcommand{\Mb}{M_\text{b}} 

\newcommand{\rGW}{\rho_\text{gw}} 
\newcommand{\OmGW}{\Omega_\text{gw}} 

\newcommand{\gmStar}{\gamma_*} 
\newcommand{\gmStarLat}{\gamma_*^\text{lat}} 

\newcommand{\phiAtMin}{\phi_\text{b}} 

\newcommand{\rVac}{\rho_\text{vac}} 
\newcommand{\OmVac}{\Omega_\text{vac}} 
\newcommand{\Vol}{{\mathcal V}} 

\newcommand{\rE}{r_\text{E}} 
\newcommand{\siTW}{\sigma^\text{tw}}
\newcommand{\phiCrit}{\phi_\text{c}}

\newcommand{\nb}{n_\text{b}} 

\definecolor{newgreen}{RGB}{10,100,20}
\definecolor{purple}{rgb}{0.5,0,0.5}
\definecolor{BLUE}{rgb}{0,0,1}

\def\lsi{\raise0.3ex\hbox{$<$\kern-0.75em\raise-1.1ex\hbox{$\sim$}}}
\newcommand{\lsim}{\mathop{\lsi}}

\graphicspath{
  {./}
  {Figures/GWPowerSpectra/}
  {Figures/GWPowerSpectra/Exponential/}
  {Figures/GWPowerSpectra/Simultaneous/}
  {Figures/GWPowerSpectra/Constant/}
  {Figures/MultiOmegaTot/}
  {Figures/MultiOmegaTot/Simultaneous/}
  {Figures/ScalarFieldEnergy/}
  {Figures/ScalarFieldEnergy/Exponential/}
  {Figures/ScalarFieldEnergy/Simultaneous/}
  {Figures/ScalarFieldEnergy/Constant/}  
}

\begin{document}

\newcommand{\Sussex}{\affiliation{
Department of Physics and Astronomy,
University of Sussex, Falmer, Brighton BN1 9QH,
U.K.}}

\newcommand{\HIPetc}{\affiliation{
Department of Physics and Helsinki Institute of Physics,
PL 64, 
FI-00014 University of Helsinki,
Finland
}}

\preprint{HIP-2018-4/TH}

\title{Gravitational waves from vacuum first-order phase transitions: \\
  from the envelope to the lattice}

\author{Daniel Cutting}
\email{d.cutting@sussex.ac.uk}
\Sussex
\HIPetc
\author{Mark Hindmarsh}
\email{m.b.hindmarsh@sussex.ac.uk}
\HIPetc
\author{David J. Weir}
\email{david.weir@helsinki.fi}
\HIPetc

\date{\today}

\begin{abstract}
We conduct large scale numerical simulations of 
gravitational wave production at 
a first-order vacuum phase transition. 
We find a power law for the gravitational wave power spectrum at
high wavenumber which falls off as $k^{-1.5}$ rather than the
$k^{-1}$ produced by the envelope approximation. 
The peak of the power spectrum is shifted to slightly lower
wave numbers from that of the envelope approximation. 
The envelope approximation 
reproduces our results for the peak power less well, agreeing only
to within an order of magnitude. 
After the bubbles finish colliding the scalar field
oscillates around the true vacuum. An additional feature
is produced in the UV of the gravitational wave
power spectrum, and this continues to
grow linearly until the end of our simulation. The additional feature
peaks at a length scale close to the bubble wall thickness and
is shown 
to have a negligible contribution to the energy in
gravitational waves, providing the scalar field
mass is much smaller than the Planck mass. 
\end{abstract}

\maketitle

\section{Introduction}

The first direct detection of gravitational waves
\cite{Abbott:2016nmj,Abbott:2016blz}
has brought in a new era of gravitational wave
astronomy. 
Future space based gravitational wave observatories such as
LISA~\cite{Audley:2017drz} hold great promise for
cosmology~\cite{Caprini:2018mtu}.
LISA's planned sensitivity band peaks at
lower frequencies than ground based detectors. It 
therefore will 
have much greater sensitivity to gravitational waves originating from
process in the very early universe. Cosmological first-order phase
transitions are one such process, and LISA's sensitivity window allows
it to probe electroweak phase transitions in many extensions of the
Standard Model~\cite{Caprini:2015zlo,Weir:2017wfa}.

In a cosmological first-order phase transition, the universe changes
from a metastable high energy (symmetric) phase to a stable lower
energy (broken) phase. This occurs through the quantum or
  thermal nucleation of bubbles of the broken phase 
  \cite{Coleman:1977py,Linde:1981zj,Steinhardt:1981ct}, 
  separated from the surrounding unbroken phase by a thin
wall. These bubbles then expand, collide and eventually coalesce. This
process generates shear stresses which in turn source gravitational
waves \cite{Witten:1984rs,1986MNRAS.218..629H}.

Early work focussed on characterising
the signal from a phase transition
that occurs in vacuum~\cite{Kosowsky:1991ua}. In such a
transition, the bubble wall quickly accelerates to ultra-relativistic
velocities.

A model of such a scenario was developed, termed the envelope
approximation~\cite{Kosowsky:1992}. In this model, the shear stresses are
assumed to be concentrated in an infinitesimally thin shell located at
the bubble wall. Upon the collision of the bubble walls, the shear stress
is assumed to dissipate, and so any regions where bubbles overlap are
ignored. The characteristic gravitational wave power spectrum from the
envelope approximation is a broken power law in wave number $k$, where
the spectrum rises as $k^3$ from the low-wavenumber (IR) direction and
falls off as $k^{-1}$ in the high-wavenumber (UV) direction. The peak
of the broken power law is associated with the length scale of the
average bubble separation $\Rstar$.

Although the envelope approximation was originally created for bubbles
expanding in vacuum it was quickly applied to thermal first-order
phase transitions, in which the scalar bubbles expand in a hot plasma
\cite{Kamionkowski:1993,Huber:2008,Konstandin:2017sat}.
In this case frictional effects from the plasma typically cause the bubble wall to approach a terminal speed $\vw$. 
which is not generally ultra-relativistic. 
Then the majority of the energy
liberated from the phase transition is deposited into heat or the bulk
motion of the plasma, and the gravitational waves sourced from the
shear stress in the scalar field are negligible.  It was argued
that, providing the shear stress in the plasma is assumed to be in
an infinitesimally thin envelope at the bubble wall, the envelope
approximation can once again be applied \cite{Kamionkowski:1993}.
Later modelling of bubble collisions introduced a thick fluid shell, and proposed that the 
velocity field should be Gaussian \cite{Caprini:2007xq}.

Large scale three-dimensional (3D) hydrodynamical 
simulations \cite{Hindmarsh:2013xza,Hindmarsh:2015qta,Hindmarsh:2017gnf}
have dramatically changed the picture.  
They show that the shear stresses do not disappear with the bubbles,  
and persist for long after the transition completes, in the form of sound waves.  The
  envelope approximation is not a good description of total
  gravitational wave production, and predicts incorrectly both the
  amplitude and shape of the gravitational wave power spectrum.
A better picture of the post-collision phase 
is one of many overlapping counter-propagating 
sound shells \cite{Hindmarsh:2016lnk}.

On the other hand, the envelope approximation
does correctly describe the sub-dominant
contribution to the power spectrum from the scalar field
\cite{Weir:2016}, 
and analytic studies within the envelope
approximation have confirmed the broken power laws found from
numerical simulations \cite{Jinno:2016vai}.
The envelope approximation can also accommodate the 
idea that fluid shells persist after collision 
\cite{Jinno:2017fby,Konstandin:2017sat}.

 It is therefore widely believed that the envelope approximation
  describes the gravitational power in cases where the energy-momentum
  of the system is dominated by the scalar field, 
  where the system is
  close to its vacuum state.  In this paper, we investigate 
the quality of the envelope approximation
  with 3D numerical simulations of a first-order vacuum phase transition.

Classical lattice simulations of a vacuum phase transition have been
used to study the power spectrum produced from bubble collisions
before~\cite{Kosowsky:1991ua,Child:2012qg}. A 1D 
  simulation of the 
  collision of two scalar field bubbles was carried out in
  Ref.~\cite{Kosowsky:1991ua}. 
and used to motivate the envelope
  approximation in Ref.~\cite{Kosowsky:1992vn}.
  
In Ref.~\cite{Child:2012qg}, it was
claimed that 
the power spectrum produced from collisions 
in 3D simulations with several bubbles 
was several orders
of magnitude smaller than that predicted by the envelope
approximation.  Furthermore, after the bubbles had finished colliding
there appeared to be an additional phase of the transition in which
the scalar field continued to oscillate around the true vacuum. During
this oscillation phase the power spectrum
continued to grow
and the peak of the spectrum shifted towards a higher frequency.  

Our numerical simulations adopt similar techniques.
However, we are able to perform simulations with many more bubbles and
higher wall velocities than Ref.~\cite{Child:2012qg}.

The simulations solve the field equations for a scalar field sourcing
gravitational waves in the linear approximation. The transition is
modelled by introducing bubbles of the broken phase as initial
conditions for the scalar field. This is done in three different ways,
modelling three different histories of bubble nucleation.
In simultaneous nucleation, we introduce all bubbles at the very start
of the simulation. In exponential nucleation simulations, we introduce
the bubbles with an exponentially increasing rate per unit volume.  In
constant nucleation, we introduce the bubbles at a constant rate.

We show power spectra for both the scalar field itself and also the
resulting gravitational wave power spectrum for all nucleation
types. We find that as we increase the wall velocity to ultra
relativistic speeds, the slope of the gravitational wave power
spectrum towards the UV becomes steeper than $k^{-1}$, and
  approaches $k^{-1.5}$.  The peak amplitude and peak location are
similar to those predicted by the envelope approximation. We provide a
fit for the power spectrum generated from bubble collisions.

We also confirm the existence of a phase after the bubble collisions have finished,
during which the scalar field oscillates around the true vacuum and
continues to source
gravitational waves. This creates an additional bump in the
power spectrum that is associated with the mass scale of the scalar
field. This continues to grow linearly until very late times, but 
we show that it
has a negligible contribution to the power spectrum in
comparison that of bubble collisions, providing that the mass of
the scalar field is smaller than the Planck mass. 

In the following section, we recap the
dynamics of the scalar field during a vacuum phase transition. This includes the
physics of the scalar field during bubble nucleation, expansion, and the eventual collision
and oscillation phases of the transition. In
Section~\ref{sec:GravWavesTheory} we describe how the scalar field
sources gravitational waves, and also describe the envelope
approximation. The numerical methods used to perform our simulations
are discussed in Section~\ref{sec:Methods}. Our results are split into
two parts; in Section~\ref{sec:ResultsScalar}, we present the
behaviour of the scalar field within our simulations, and in
Section~\ref{sec:ResultsGrav}, we analyse the gravitational wave power spectra
from our simulations and compare them to the envelope
approximation. Our conclusions are listed in Section~\ref{sec:Conclusions}.

\section{Dynamics of vacuum transitions} \label{sec:Dynamics}

\subsection{Scalar field dynamics}
In a first-order vacuum transition, bubbles of a new phase of a scalar field nucleate and then expand at ultra-relativistic speeds. At the interface between the two phases a bubble wall forms. In this region the scalar field varies smoothly between the two vacuum expectation values. Upon the collision and subsequently merger of the bubbles the shear stress of the system will source gravitational waves. 
The shear stress in a vacuum transition is predominantly due to gradients in the scalar field $\phi$. 

In this work, we 
study transitions in which the duration of the phase transition is much shorter than the Hubble time $H_*^{-1}$ when the transition takes place. 
For such transitions the expansion of the universe can be neglected, and the equation of motion for the scalar field is simply given by
\begin{equation}
\label{e:PhiEqu}
\Box \phi - V'(\phi)=0\text,
\end{equation}
where $V(\phi)$ is the effective potential of the scalar field.
This is sufficient to investigate the envelope approximation, but may not give accurate results for 
transitions in which the Universe enters an inflationary phase before bubbles start nucleating.

For these purposes it is sufficient to adopt a simple quartic form for the effective potential, 
\begin{equation}
V(\phi)= \frac{1}{2} M^2 \phi^2 +\frac{1}{3} \delta \phi^3 +\frac{1}{4} \lambda \phi^4\text,
\end{equation}
where the presence of a cubic term allows us to ensure the transition is first-order. The value of the scalar field in the broken phase is then
\begin{equation}
\phiAtMin=\frac{-\delta+\sqrt{\delta^2-4M^2\lambda}}{2\lambda}\text,
\end{equation}
with mass
\begin{equation}
\Mb=\sqrt{-\delta \phiAtMin -2M^2}\text,
\end{equation}
The potential difference between the two minima is given by 
\begin{equation}
\rVac= \frac{1}{12 \lambda} \left(\Mb^4- M^4\right)\text.
\end{equation}

By varying the couplings $M^2$, $\delta$ and $\lambda$ we are able to change the potential difference $\rVac$ between the two minima of our potential, and also the height of the potential barrier.

The total energy density in the scalar field $\rho_\phi$ can be split into three components,
\begin{equation}
\rho_\phi=\rho_\text{K}+\rho_{V}+\rho_\text{D}\text,
\end{equation}  
with the kinetic energy density,
\begin{equation}
\rho_\text{K}=\frac{1}{2}\dot{\phi}^2\text,
\end{equation} 
the gradient energy density,
\begin{equation} 
\rho_\text{D}=\frac{1}{2} (\nabla \phi)^2\text,
\end{equation}
and the potential energy density,
\begin{equation}
\rho_{V}=V(\phi)-V(\phiAtMin)\text.
\end{equation}

\subsection{Bubble nucleation}

In a first-order vacuum transition, bubbles nucleate by quantum
tunnelling through a potential barrier. This means that they nucleate
as critical bubbles, $\mathrm{O}(4)$-symmetric solutions to the
Euclidean field equations \cite{Coleman:1977py,Linde:1981zj}.  When
the radius of the critical bubble is much larger than the thickness of
the bubble wall the bubble is said to be in the thin wall limit. This
occurs when $\rVac$ is much smaller than the height of the potential 
barrier, or equivalently 
when the minima are close to degenerate. For our potential the minima are degenerate for 
\begin{equation}
\delta= - \frac{3}{\sqrt{2}}M\sqrt{\lambda}\text.
\end{equation}
When $\rVac$ is much larger than the height of the potential barrier, 
the critical bubble is of a similar size to the radius. 
We leave the study of such bubbles to a later work. 

The thin wall 
solution can be calculated analytically as a function of Euclidean
radius, $\rE = \sqrt{r^2 + \ta^2}$, where $r$ is the spatial radius,
and $\ta$ the Euclidean time.  In the thin wall limit the scalar field
profile of the critical bubble is given by
\begin{equation}
\label{eq:tw_profile}
\phiCrit(r)=\frac{\phiAtMin}{2}\left[1-\text{tanh}\left(\frac{r-\Rc^\text{tw} }{l_0^\text{tw}}\right)\right]\text,
\end{equation}
where $l_0$ is thickness of the bubble wall, which is given in the
thin wall limit by
\begin{equation}
l_0^\text{tw}=\frac{2}{\sqrt{V''(\phi_b)}}\text,
\end{equation}
and $\Rc ^\mathrm{tw}$ is the radius of the critical bubble.
The radius of the critical bubble 
can be estimated by 
extremising the approximate expression for the Euclidean action 
\begin{equation}
S_4 = 2\pi^2 R^3 \siTW  -  \frac{\pi^2}{2}  R^4 \rVac \text,
\end{equation}
where 
\begin{equation}
\siTW = \frac{M^3}{3\lambda}\text,
\end{equation}
is interpreted as the surface tension of the bubble. 
Then the critical radius is 
\begin{equation}
\Rc ^\mathrm{tw}=\frac{3 \sigma^\text{tw}}{\rVac}\text.
\end{equation}
At the point of time symmetry $\ta = 0$, 
the energy liberated from the vacuum is equal to the energy in the wall.
Furthermore, the outward force on the bubble wall due to the pressure difference $\rVac$ 
is equal and opposite to that caused by the surface tension. 

Once the bubble has nucleated,
the solution is found by the analytic continuation to Minkowski space, so that
\begin{equation}
\phi(r,t) = \left\{ 
\ba{cc}
\phiCrit (\sqrt{r^2 - t^2}), & r > t\text, \\
\phiAtMin, & r \le t \text.
\ea
\right.
\end{equation}

The probability of nucleating a bubble per unit volume per unit time $p(t)$ is given by \cite{Linde:1981zj}
\begin{equation}
p(t)=p_\text{n}\exp(-S_4)\text.
\end{equation}
Very often the Euclidean action decreases slowly in time due to a
change in temperature or a background field.
Then we may write 
\begin{equation}\label{eq:exp_nuc}
p(t)=p_f\exp[\beta(t-t_f)]\text,
\end{equation}
where $\be = - \left. d \ln p(t) /d t \right|_{t_f} $ and $t_f$ is the time at which the fraction
of the universe in the symmetric phase is $h(t_f)=1/e$ \cite{PhysRevD.45.3415}. 
The bubble number density can be shown to be 
\ben
\label{e:BubDenExp}
\nb = \frac{1}{8\pi} \frac{\be^3}{\vw^3},
\een 
where in the vacuum case $\vw$ is very close to unity.
We will refer to this case as exponential nucleation.

It is also possible that $S_4(t)$ has a minimum which is reached at
time $t_0$ before a transition completes.  Then the probability of
nucleating a bubble per unit volume could be
approximated by
\begin{equation} \label{eq:sim_nuc}
p(t)=p_0\exp[{- \textstyle\half}\beta^2_2 (t-t_0)^2 ]\text,
\end{equation}
where $\beta_2= \sqrt{S''(t_0)}$. 
Nucleation is then concentrated around time $t_0$ \cite{Jinno:2017ixd}. 
The bubble density is 
\ben
\label{e:BubDenSim}
\nb =  \sqrt{2\pi}\frac{p_0}{\be_2}.
\een 
We will refer to this case as 
simultaneous nucleation.

The last possibility we consider is if $S_4(t)$ tends to a
constant (for a model with a constant nucleation rate see Ref.~\cite{GarciaGarcia:2016xgv}). 
We would then expect bubbles to nucleate at a constant rate
\begin{equation}
p(t)=p_\text{c},
\end{equation}
for which 
\ben
\label{e:BubDenCon}
\nb = \frac{1}{4} \left(\frac{3}{\pi }\right)^{1/4} \Gamma\left(\frac{1}{4}\right) \left(\frac{p_\text{c}}{v_\text{w}}\right)^{3/4} . 
\een
We will refer to this as constant nucleation.

\subsection{Bubble growth}

If we consider a thin wall bubble then we can obtain an expression for
the evolution of the bubble simply by considering energy
conservation. The energy in the static bubble wall per unit area is
simply $\siTW$.
Then if the bubble wall is expanding at some velocity $\vw$, the
energy per unit area is given by $\sigma^\text{tw} \gamma$ where
$\gamma $ is the wall's Lorentz factor.  The total energy of an
  expanding bubble with radius $R$ is then \cite{Coleman:1977py,Linde:1981zj}
\begin{equation}
E_\mathrm{bub}=4\pi R^2 \sigma^\text{tw} \gamma -\frac{4}{3}\pi R^3 \rVac,
\end{equation}
where we can define $R$ to be the point in the scalar field profile such that $\phi(R)=\phi_b/2$. 
As we are considering vacuum decay,
we expect $E_{\rm bub}=0$. We therefore obtain that, for a bubble of radius $R$, the Lorentz factor of the bubble wall is given by
\begin{equation}
\label{eq:gamma_bubble_growth}
 \gamma(R)=\frac{R\rVac}{3\sigma^\text{tw}}=\frac{R}{\Rc ^\mathrm{tw}}\text. 
\end{equation}
We expect Eq.~(\ref{eq:gamma_bubble_growth}) to apply outside the thin wall limit by recalling that the solution of the classical field equations is simply the analytic continuation of the O(4)-symmetric bounce solution \cite{Coleman:1977py}. Then any point in the field profile of the critical bubble we define to be the critical radius $\Rc $ will expand out with a hyperboloid motion satisfying
\begin{equation}
R^2(t)-t^2=\Rc ^2\text,
\end{equation} 
which is equivalent to Eq.~(\ref{eq:gamma_bubble_growth}).

\subsection{Bubble collision and oscillation phase}\label{sec:CollOsc}

\begin{figure}
\centering 
\includegraphics[width=0.48\textwidth]{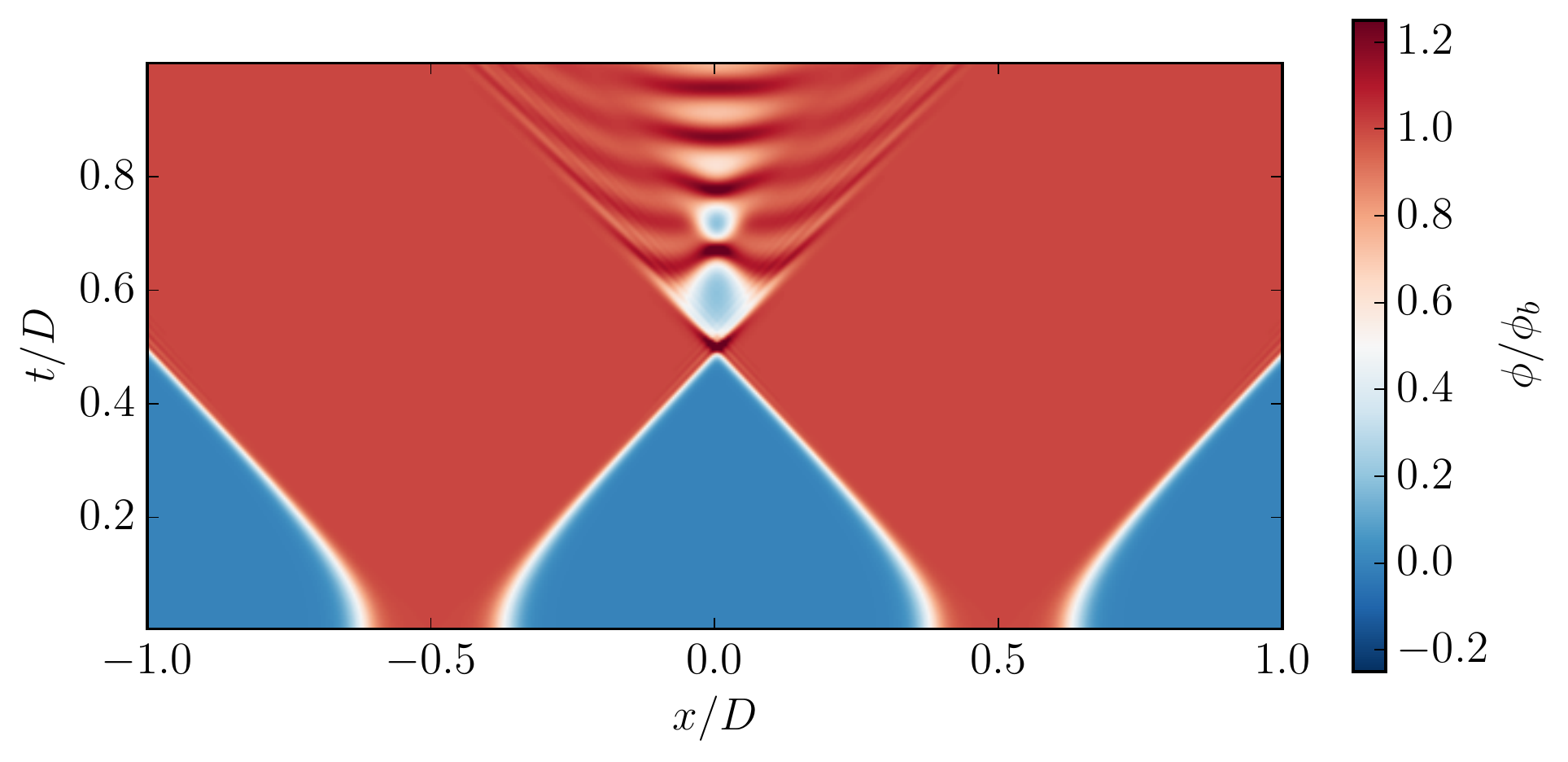}
\caption{Values of the scalar field along the collision axis during a
  two bubble collision where $\Rc M=7.15$. Here the $x$ axis is the
  collision axis which connects the two bubble centres. The $y$ axis
  is time since the nucleation of the bubbles. The bubbles are separated by a distance $D$.  This figure can be
  compared with Fig.~1 of \cite{PhysRevD.26.2681} and Fig.~7 of
  \cite{Braden:2014cra}. }
\label{fig:CollAxis}
\end{figure}

For bubbles with thin walls, 
after collision part of the overlap region 
rebounds and returns 
towards the false vacuum~\cite{PhysRevD.26.2681,Braden:2014cra}. 
In Fig.~\ref{fig:CollAxis}, we plot the variation during a collision of the scalar
field along the collision axis connecting two bubble centres. At the collision point it can be seen that the
scalar field oscillates between the true and false vacuum. 
These large amplitude oscillations are the  
source of scalar radiation moving at close to the speed of
light, and can also induce rapid production of light particles through parametric resonance \cite{Zhang:2010qg}.
This rebounding and oscillation phase is something that is not
accounted for within the envelope approximation. Away from the thin wall limit, 
the scalar field in the overlap region is not able to return to the false vacuum, and instead will just oscillate around the true vacuum 
\cite{Braden:2014cra}. 

After this stag, the scalar field continues to oscillate around the
true vacuum with large amplitude
oscillations.  In the absence of other interactions, scalar fields take
a substantial time to thermalise~\cite{Aarts:2000mg,Micha:2002ey,Arrizabalaga:2005tf}.

\section{Gravitational waves from a phase transition} \label{sec:GravWavesTheory}

In order to calculate 
the gravitational wave power spectrum, we need to find the transverse traceless (TT) metric perturbations $h^{TT}_{ij}$ where 
\begin{equation}
\Box h^{TT}_{ij} = 16\pi G T^{TT}_{ij}\text,
\end{equation}
and $T^{TT}_{ij}$ is the transverse traceless projection of the energy-momentum tensor,
\begin{equation}
T_{\mu\nu} = \partial_\mu\phi \partial_\nu\phi - \eta_{\mu\nu} \left(  \half (\partial\phi)^2 + V(\phi) \right), 
\end{equation}
where $\eta_{\mu\nu}$ is the Minkowski metric. 
The energy density in the gravitational waves can be defined as 
\begin{equation}
\label{e:GWEnDen}
\rGW(\bx,t) = \frac{1}{32\pi G}  \dot h^{TT}_{ij}\dot h^{TT}_{ij}. 
\end{equation}
Note that an average over many wavelengths and periods may be needed in order to reduce fluctuations in this quantity. 

We introduce an auxiliary tensor $u_{ij}$ which satisfies 
\cite{GarciaBellido:2007af}
\begin{equation}
\label{e:uEqu}
\Box u_{ij} = 16\pi G (\partial_i \phi)( \partial_j \phi)\text.
\end{equation}
To obtain $h^{TT}_{ij}$ we use the projector $\Lambda_{ij,lm}$ on $u_{ij}$ in momentum space,
\begin{equation}\label{eq:ProjMetric}
    h^{TT}_{ij}(\mathbf{k},t) = \Lambda_{ij,lm}(\mathbf{k})u_{lm}(\mathbf{k},t)\text,
\end{equation}
where
\begin{equation}    
\Lambda_{ij,lm}(\mathbf{k})=P_{im}(\mathbf{k})P_{jl}(\mathbf{k})-\frac{1}{2}P_{ij}(\mathbf{k})P_{lm}(\mathbf{k})\text,
\end{equation}
and
\begin{equation}    
 P_{ij}(\mathbf{k})=\delta_{ij}-\hat{k}_i\hat{k}_j\text.
\end{equation}
We then define the spectral density of the time derivative of the metric perturbations $P_{\dot{h}}$ as
\begin{equation}
\langle \dot{h}^{TT}_{ij}(\mathbf{k},t) \dot{h}^{TT}_{ij}(\mathbf{k'},t) \rangle =P_{\dot{h}}(\mathbf{k},t) (2\pi)^3 \delta(\mathbf{k}+\mathbf{k'})\text.
\end{equation}
Therefore the power spectrum of gravitational wave energy density is
\begin{equation}
\frac{d\rGW}{d\mathrm{ln}(k)}=\frac{1}{32\pi G} \frac{k^3}{2\pi^2}P_{\dot{h}}(\mathbf{k},t)\text,
\end{equation}
and by dividing through by the critical energy density $\rho_c$ we obtain the power spectrum of the gravitational wave energy density parameter
\begin{equation}
\frac{d\OmGW}{d\mathrm{ln}(k)}=\frac{1}{32\pi G \rho_c} \frac{k^3}{2\pi^2}P_{\dot{h}}(\mathbf{k},t)\text.
\end{equation}

\subsection{Collision Phase: envelope approximation}

In the envelope approximation~\cite{Kosowsky:1992}, 
the bubble walls are treated as infinitely thin, expanding with speed $\vw$, and containing all the vacuum energy released by the transition.  
The overlap region of collided bubbles are ignored, and the 
gravitational waves from shear stress ``envelope'' calculated. 
The resulting spectrum was re-computed with many more bubbles in Ref.~\cite{Huber:2008}, and again 
in Ref. \cite{Konstandin:2017sat}, for an exponential nucleation rate in both cases.

The gravitational wave power spectrum is well approximated by a 
broken power law 
\begin{equation}
\frac{d\OmGW^\text{env}}{d\mathrm{ln}(k)}= {\Om}^\text{env}_\text{p} \frac{(a+b)\tilde{k}^bk^a}{b\tilde{k}^{(a+b)}+ak^{(a+b)}}\text,
\end{equation}
with power law exponents $a$ and $b$, peak amplitude ${\Om}^\text{env}_\text{p}$ and peak wavenumber $\tilde{k}$. 
The peak amplitude was found to be 
\begin{equation}
{\Om}^\text{env}_\text{p}\simeq \frac{0.44v_w^3}{1+8.28v_w^3}\left(\frac{H_*}{\beta}\right)^2 \left(\kappa_\phi \Om_\mathrm{vac}\right)^2\text,
\end{equation}
where the Hubble rate at the time of the transition $H_*$, the vacuum energy density parameter $\Om_\mathrm{vac}=\rVac/\rho_c$ and the bubble wall velocity $\vw$. 
The peak frequency was estimated to be
\begin{equation}
\tilde{k}/\beta\simeq \frac{1.96}{1-0.051\vw+0.88\vw^2}\text.
\end{equation}

The efficiency factor $\kappa_\phi$ measures the fraction of vacuum energy that is converted to stress energy localised at the bubble wall. 
We define it as
\begin{equation}
\kappa_\phi=\frac{2 \rho_D}{\rVac-\rho_\text{V}}\text.
\end{equation}
For a vacuum phase transition $\kappa_\phi\simeq1$.

The exponent for the broken power law on the low frequency side is expected to be $a = 3$ due to causality \cite{Caprini:2009fx}. 
In Ref.~\cite{Konstandin:2017sat}, the power law exponents were found to be $a= 2.9$ and $b= 0.9$ for $\vw\simeq1$ and $a= 2.95$ and $b= 1$ for $\vw\ll 1$. 
Furthermore, in Ref.~\cite{Weir:2016}, the envelope approximation was compared to lattice simulations of a scalar field with
frictional effects chosen such that the bubble walls asymptotes a constant speed $\vw=0.44$. The gravitational wave power spectrum generated by stress energy in the scalar field was found to agree well with the envelope approximation. The power law exponents for the envelope approximation in this study were found to be $a =2.98 \pm 0.02$ and $b =0.62 \pm 0.05$ \cite{Weir:2016}.  

There is also some analytical understanding of the power spectrum produced under the envelope approximation. In Ref. \cite{Jinno:2016vai}, it is shown that the two point correlator of the energy-momentum tensor can be expressed 
as a 1-dimensional integral
under the envelope approximation, also producing 
a broken power law with exponents $a=3$ and $b=1$.

It should be noted that while typically in a thermal phase transition friction effects from the plasma cause $\vw$ to approach a constant, in a vacuum phase transition the bubble wall accelerates until collision with $\vw \rightarrow 1$ and $\gamma \rightarrow \infty$.
In this current work we shall check 
whether the formula is a good fit in the case 
where the bubble wall continues to accelerate until collision, reaching ultra-relativistic velocities.

\subsection{Oscillation phase}

Previous simulations of a vacuum first-order phase transition have
observed that after all the bubble collisions have completed, 
the scalar field continues to oscillate, and
the production of gravitational
radiation continues \cite{Child:2012qg}. 

The contribution to the gravitational wave power spectrum from this 
oscillation phase was seen to dominate that of the bubble collisions by more than
an order of magnitude. The peak frequency moved towards the UV
by an order of magnitude during the oscillation phase. 

Providing the oscillations in the scalar field are non-linear \cite{Dufaux:2007pt} we would
expect them to be a continuous source of
gravitational waves, similar to acoustic waves in a thermal phase
transition \cite{Hindmarsh:2015qta}. Eventually Hubble friction would
damp out the oscillations. 

A further goal of the current work is to investigate if we also see 
the growth of the gravitational wave power spectrum during an oscillation phase in our simulations.

\section{Methods}\label{sec:Methods}

To perform our study we perform a series of simulations 
solving the partial differential equations (\ref{e:PhiEqu}) and (\ref{e:uEqu}) on 
on a 3-dimensional lattice, using code built on 
the open source \texttt{C++} library \texttt{LAT}field2 \cite{David:2015eya}. 
To compute derivatives we use a central finite difference method. For the Laplacian we use the minimal 7-point stencil made up of a central point and then an additional 2 points in each dimension. We choose our timestep $\Delta t$ and lattice spacing $\Delta x$ such that $\Delta t= 0.2\Delta x$. We advance in timestep by using the leapfrog algorithm.

Our simulations are on a cubic grid with total volume $\Vol=(L \Delta x)^3$, and periodic boundary conditions. We begin each simulation by nucleating at least one bubble at the simulation time $t=0$. The total number of bubbles nucleated by the end of the simulation is given by $\Nb$. 

We use a shooting method to find the critical profile for specific values of $M^2$, $\delta$ and $\lambda$. We choose three profiles to simulate and give the parameters for these in Table~\ref{table:potential}. The resulting field profiles are modelled well by Eq.~(\ref{eq:tw_profile}) 
with values for $\Rc$ and $l_0$ given in the first two columns. 
Note that they differ from the thin wall values due to the finite size of the bubble.  

The value of $\Rc$ is given by the location in the numerical profile at which 
\begin{equation}
\phi_c(\Rc)=\frac{\phiAtMin}{2}\text.
\end{equation}
Similarly, $l_0=r^+-r^-$ where 
\begin{equation}
\phi_c(r^\pm)=\frac{\phiAtMin}{2}\left(1-\tanh\left(\pm 1/2\right)\right)\text.
\end{equation}

\begin{table}
\centering

\begin{tabular}{l c | c  c | c c |c c r }
\multicolumn{1}{c}{$\Rc  M$} & \multicolumn{1}{c|}{$l_0 M$} & \multicolumn{1}{c}{$\delta/M$}  & \multicolumn{1}{c|}{$\lambda$} & \multicolumn{1}{c}{$\phiAtMin/M$} & \multicolumn{1}{c|}{$\rVac/M^4$}             & \multicolumn{1}{c}{$\sigma^\mathrm{tw}/M^3$}  & \multicolumn{1}{c}{$l_0^\mathrm{tw} M$}   & \multicolumn{1}{c}{$\Rc ^\mathrm{tw} M$}  \\ \hline
7.15     &  1.71   & -1.632      & 0.5       & 2.45          & 0.495                   & 2/3                       & 1.42                  & 4.04                  \\
14.3     & 1.83    & -1.56       & 0.5       & 2.22          & 0.189                   & 2/3                       & 1.65                  & 10.6                  \\
28.8     & 1.91    & -1.528      & 0.5       & 2.11          & 0.0809                  & 2/3                       & 1.81                  & 24.7                  \\ 
\end{tabular}
\caption{Critical radii $\Rc$ and wall thicknesses $l_0$ that are used in our simulations. For each of these we give the potential parameters $\delta$ and $\lambda$ used to derive them, the broken phase value of the scalar field $\phiAtMin$ and the vacuum energy density $\rVac$. We also list the surface tension $\siTW$, wall thickness $l_0^\mathrm{tw}$ and critical radius $\Rc^\mathrm{tw}$ as derived from the thin wall approximation.  }\label{table:potential}

\end{table}

We nucleate bubbles with a critical profile inside our numerical simulation. Before nucleating the $N$th bubble we check that, for all $n<N$, the distance between the $N$th and $n$th bubble centres $r_n$ obeys the following relation
\begin{equation}
r_n^\text{sep}>\Rc+\sqrt{\Rc^2+(t-t_n)^2}\text,
\end{equation}
where $t_n$ is the time at which the $n$th bubble nucleated. Providing this is satisfied for all bubbles, we nucleate a bubble by modifying $\phi\rightarrow \phi'$, where
\begin{equation}
\phi'(r)=\sqrt{\phi^2(r) + \phi_c^2(r)}\text.
\end{equation}

We evolve the auxiliary metric tensor $u_{ij}$ in real space at every timestep. At routine intervals we perform a Fourier transform of $\dot{u}_{ij}$, and then project the result according to Eq.~(\ref{eq:ProjMetric}) to find $\dot{h}^{TT}_{ij}(\mathbf{k},t)$. From this we then calculate the gravitational wave power spectrum. It should be noted that in our units $G=1$, though in general we plot quantities that do not depend on G.

We can nucleate bubbles simultaneously at the start of the simulation, or indeed with a nucleation rate throughout its duration. 
In order to compare with earlier studies using the envelope approximation, we nucleate bubbles with an exponentially increasing nucleation rate, 
using the algorithm given in Ref.~\cite{PhysRevD.45.3415}. 
Then the probability of nucleating a bubble per unit volume and time is given by Eq.~(\ref{eq:exp_nuc}). 
The parameters of the simultaneous nucleation runs are listed in Table~\ref{table:sim}, 
and those of the exponential nucleation runs in
Table~\ref{table:exp}. We also perform two constant nucleation runs to
check that this type of nucleation is consistent with our other
results. The parameters of the constant nucleation runs are given in Table~\ref{table:con}. 

We also wish to study the gravitational wave power spectrum produced after the bubble collision phase is completed. In order to do this, we can simply turn on the evolution of $u_{ij}$ once the bubbles have finished colliding. We employ this approach for a series of simultaneous nucleation simulations listed in Table~\ref{table:late-sim}.

\begin{table}

\centering
\begin{tabular}{l c c |r c r c | r}
\multicolumn{1}{c}{$\Rc M$}& \multicolumn{1}{c}{$\gmStar$}  &\multicolumn{1}{c|}{$\Rstar M$}   & \multicolumn{1}{c}{$N_b$} & \multicolumn{1}{c}{$L\Delta x M$}  & \multicolumn{1}{c}{$L$}    & \multicolumn{1}{c|}{$\Delta xM$}&  \multicolumn{1}{c}{$\gmStarLat$}\\ \hline
7.15   &   1.97     &    28.2     & 8     & 56.32          & $128$  & 0.44       & 1.85         \\
7.15   &   1.97     &    28.2     & 64    & 112.64         & $256$  & 0.44       & 1.85         \\
7.15   &   1.97     &    28.2     & 512   & 225.28         & $512$  & 0.44       & 1.85         \\
7.15   &   1.97     &    28.2     & 4096  & 450.56         & $1024$ & 0.44       & 1.85         \\ \hline
7.15   &   3.94     &    56.3     & 8     & 112.64         & $512$  & 0.22       & 3.37         \\
7.15   &   3.94     &    56.3     & 64    & 225.28         & $1024$ & 0.22       & 3.37         \\
7.15   &   3.94     &    56.3     & 512   & 450.56         & $2048$ & 0.22       & 3.37         \\
7.15   &   3.94     &    56.3     & 4096  & 901.12         & $4096$ & 0.22       & 3.37         \\ \hline
7.15   &  7.88      &    113.     & 8     & 225.28         & $2048$ & 0.11       & 5.65         \\ \hline
14.3   &  1.97      &    56.3     & 512   & 450.56         & $1024$ & 0.44       & 1.87         \\ 
28.8   &  1.96      &    113.     & 512   & 901.12         & $2048$ & 0.44       & 1.89         \\
\end{tabular}                                      
\caption{Parameters of the simultaneous nucleation simulations used within this paper. Listed here for each run is the critical radius $\Rc$, typical Lorentz factor at collision $\gmStar$, average bubble seperation $\Rstar$, number of bubbles $\Nb$, number of lattice points $L^3$, lattice spacing $\Delta x$, and effective $\gmStar$ as found on the lattice $\gmStarLat$. Not given here are simulation runs where the metric perturbations are  turned on after the bubbles have finished colliding, see Table~\ref{table:late-sim}.
}\label{table:sim}

\end{table}

\begin{table}

\centering
\begin{tabular}{l c c c | r c  r c | r}

\multicolumn{1}{c}{$\Rc M$}& \multicolumn{1}{c}{$\gmStar$}  &\multicolumn{1}{c}{$\beta/M$}& \multicolumn{1}{c|}{$\Rstar M$}& \multicolumn{1}{c}{$N_b$} & \multicolumn{1}{c}{$L \Delta x M$}  &  \multicolumn{1}{c}{$L$}  & \multicolumn{1}{c|}{$\Delta xM$}& \multicolumn{1}{c}{$\gmStarLat$}\\ \hline
7.15   & 1.97       & 0.180   & 28.2      & 8     & 56.32           & $128$ & 0.44       & 1.85        \\
7.15   & 1.92       & 0.180   & 27.5      & 69    & 112.64          & $256$ & 0.44       & 1.81        \\
7.15   & 1.96       & 0.180   & 28.0      & 522   & 225.28          & $512$ & 0.44       & 1.84        \\
7.15   & 3.94       & 0.0625  & 56.3      & 8     & 112.64          & $512$ & 0.22       & 3.37        \\
7.15   & 4.09       & 0.0625  & 58.5      & 57    & 225.28          & $1024$& 0.22       & 3.55        \\
7.15   & 7.57       & 0.0290  & 108.      & 9     & 225.28          & $2048$& 0.11       & 5.58        \\
\end{tabular}
\caption{Parameters of the exponential nucleation simulations used within this paper.}
\label{table:exp}

\end{table}

\begin{table}

\centering
\begin{tabular}{l c c c | r c  r c | r}

\multicolumn{1}{c}{$\Rc M$}& \multicolumn{1}{c}{$\gmStar$}  &\multicolumn{1}{c}{$p_\mathrm{c}/M^4$}    & \multicolumn{1}{c|}{$\Rstar M$}& \multicolumn{1}{c}{$N_b$} & \multicolumn{1}{c}{$L \Delta x M$}  &  \multicolumn{1}{c}{$L$}  & \multicolumn{1}{c|}{$\Delta xM$}& \multicolumn{1}{c}{$\gmStarLat$}\\ \hline
7.15   & 3.94       &$1.50\times 10^{-7}$  & 56.3     & 64     & 225.28          & $1024$& 0.22       & 3.37        \\
7.15   & 4.09       &$1.50\times 10^{-7}$  & 56.3     & 510    & 450.56          & $2048$& 0.22       & 3.37        \\
\end{tabular}
\caption{Parameters of the constant nucleation simulations used within this paper.}
\label{table:con}

\end{table}

\begin{table}

\centering

\begin{tabular}{l c c |r c r c| r}
\multicolumn{1}{c}{$\Rc  M$} & \multicolumn{1}{c}{$\gmStar$} &  \multicolumn{1}{c|}{$\Rstar  M$} & \multicolumn{1}{c}{$N_b$} & \multicolumn{1}{c}{$L \Delta x M$}  &  \multicolumn{1}{c}{$L$}  & \multicolumn{1}{c|}{$\Delta x M$} &\multicolumn{1}{c}{$\gmStarLat$}\\ \hline
7.15     & 3.94      &  56.3        &  64   & 225.28          & $1024$& 0.44         & 3.37       \\
\hline
14.3     & 1.97      &  56.3        &  512  & 450.56          & $1024$& 0.44         & 1.87       \\
14.3     & 3.94      &  113.        &  8    & 225.28          & $1024$& 0.22         & 3.34       \\\hline
28.8     & 1.96      &  113.        &  64   & 450.56          & $1024$& 0.44         & 1.89       \\
\end{tabular}
\caption{Parameters of the simultaneous nucleation simulations where the metric perturbations are turned on at $t/\Rstar=2.0$ at which point most of the bubbles have finished colliding. This is in order to see the shape of the power spectrum due to scalar field radiation during the oscillation phase.}\label{table:late-sim}
\end{table}

\begin{figure*}
\subfigure[]{\includegraphics[width=0.485\textwidth,clip]{{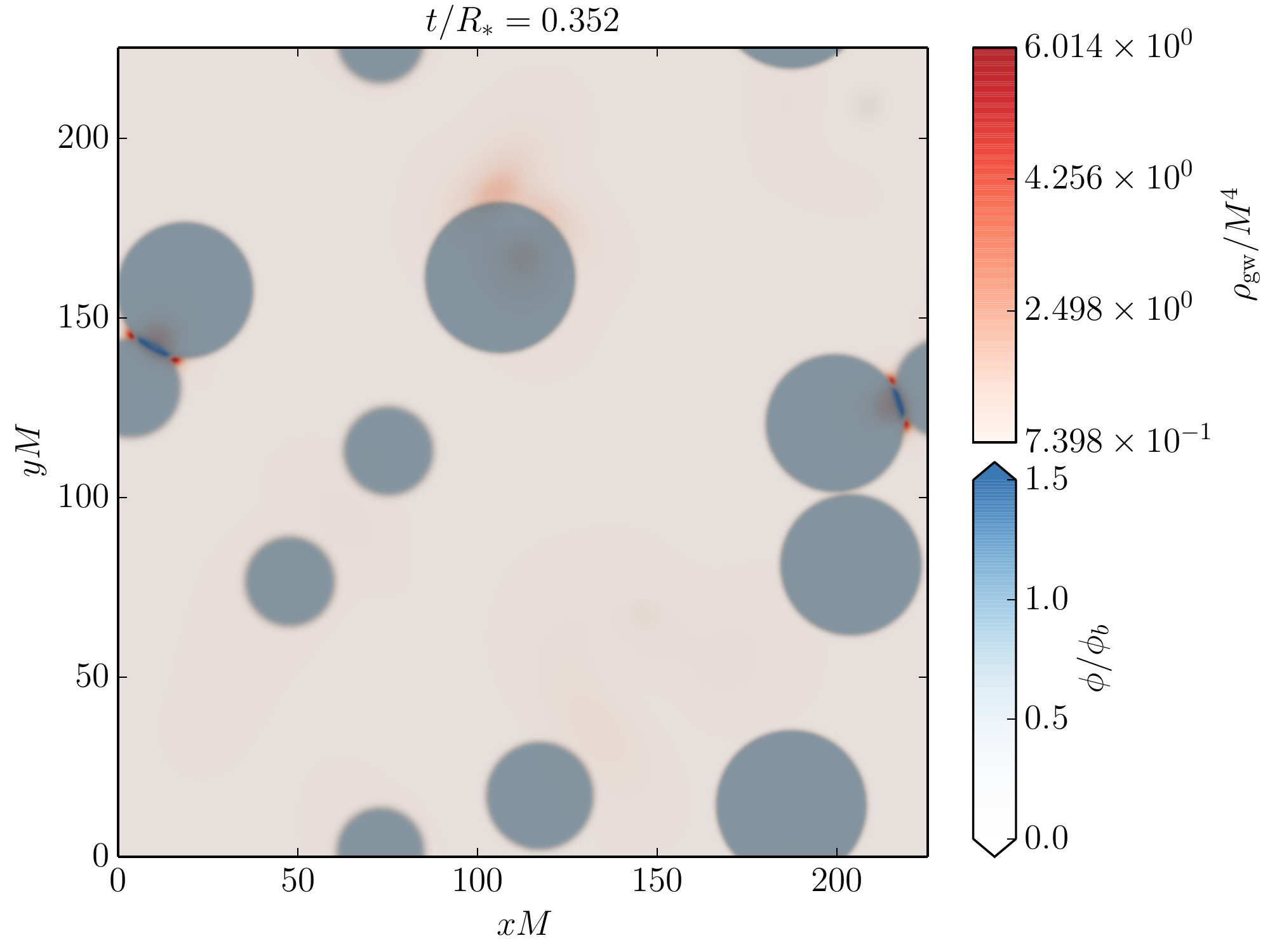}}}
\hfill
\subfigure[]{\includegraphics[width=0.485\textwidth,clip]{{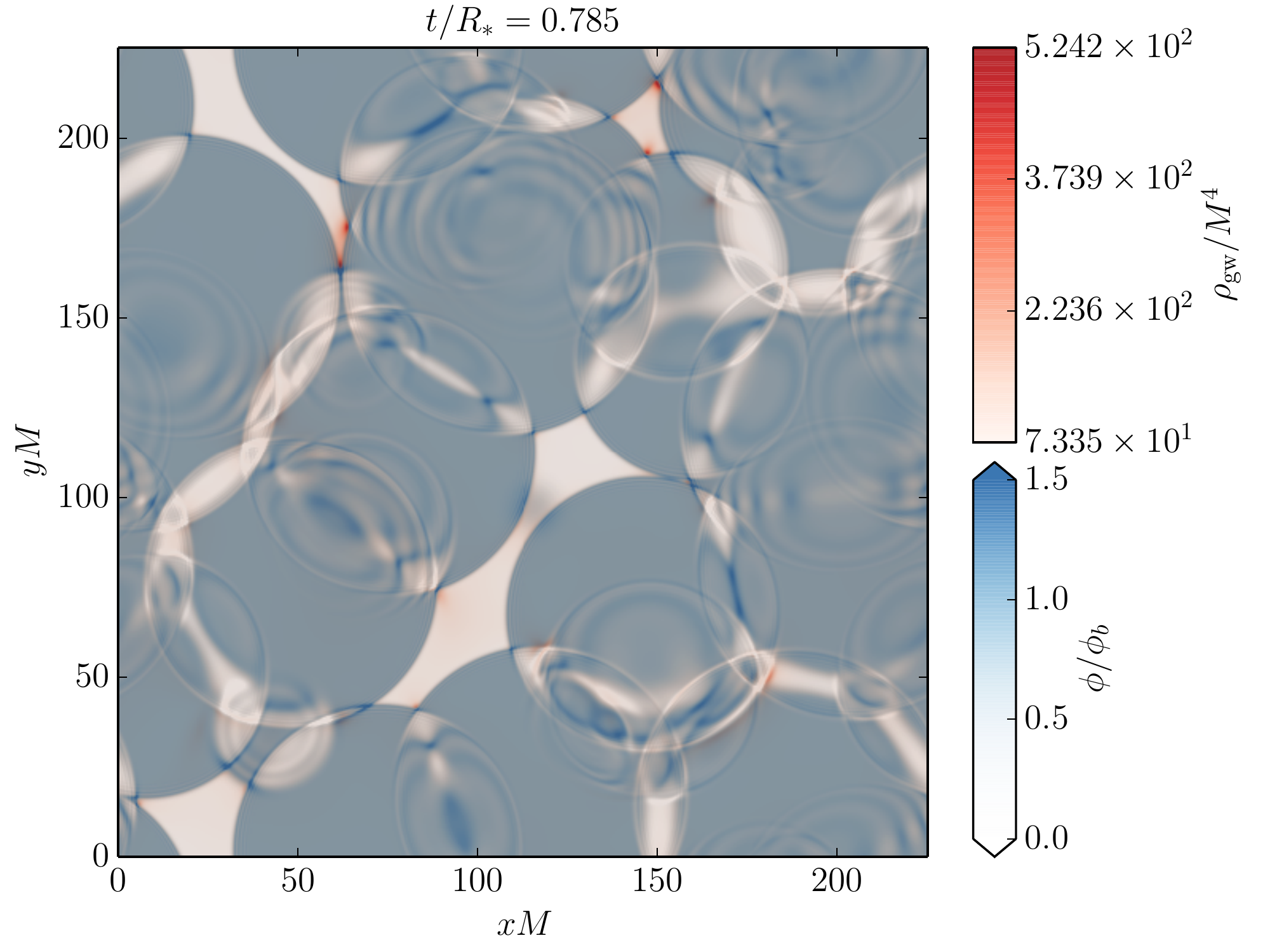}}}
\\
\subfigure[]{\includegraphics[width=0.485\textwidth,clip]{{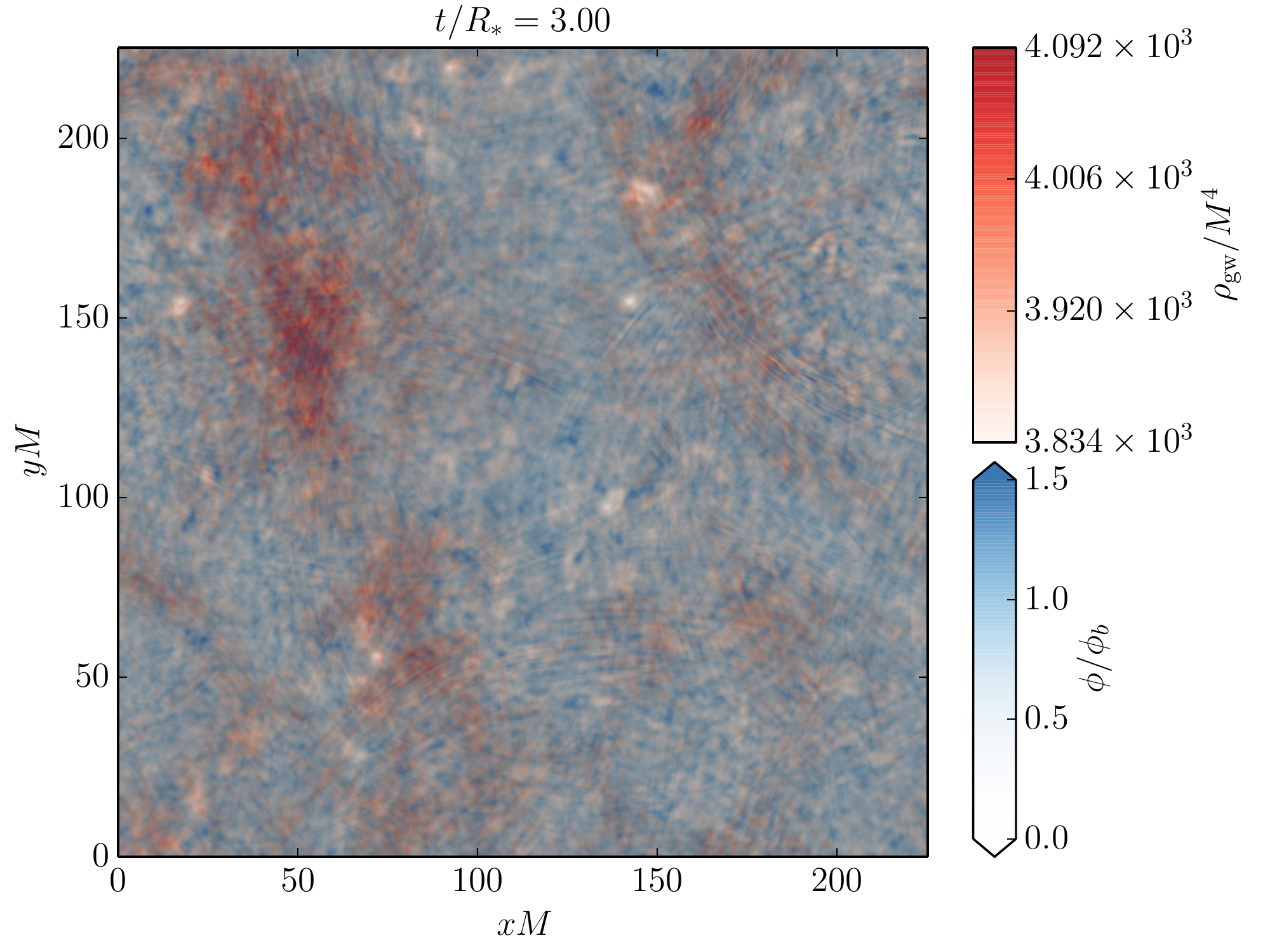}}}
\hfill
\subfigure[]{\includegraphics[width=0.485\textwidth,clip]{{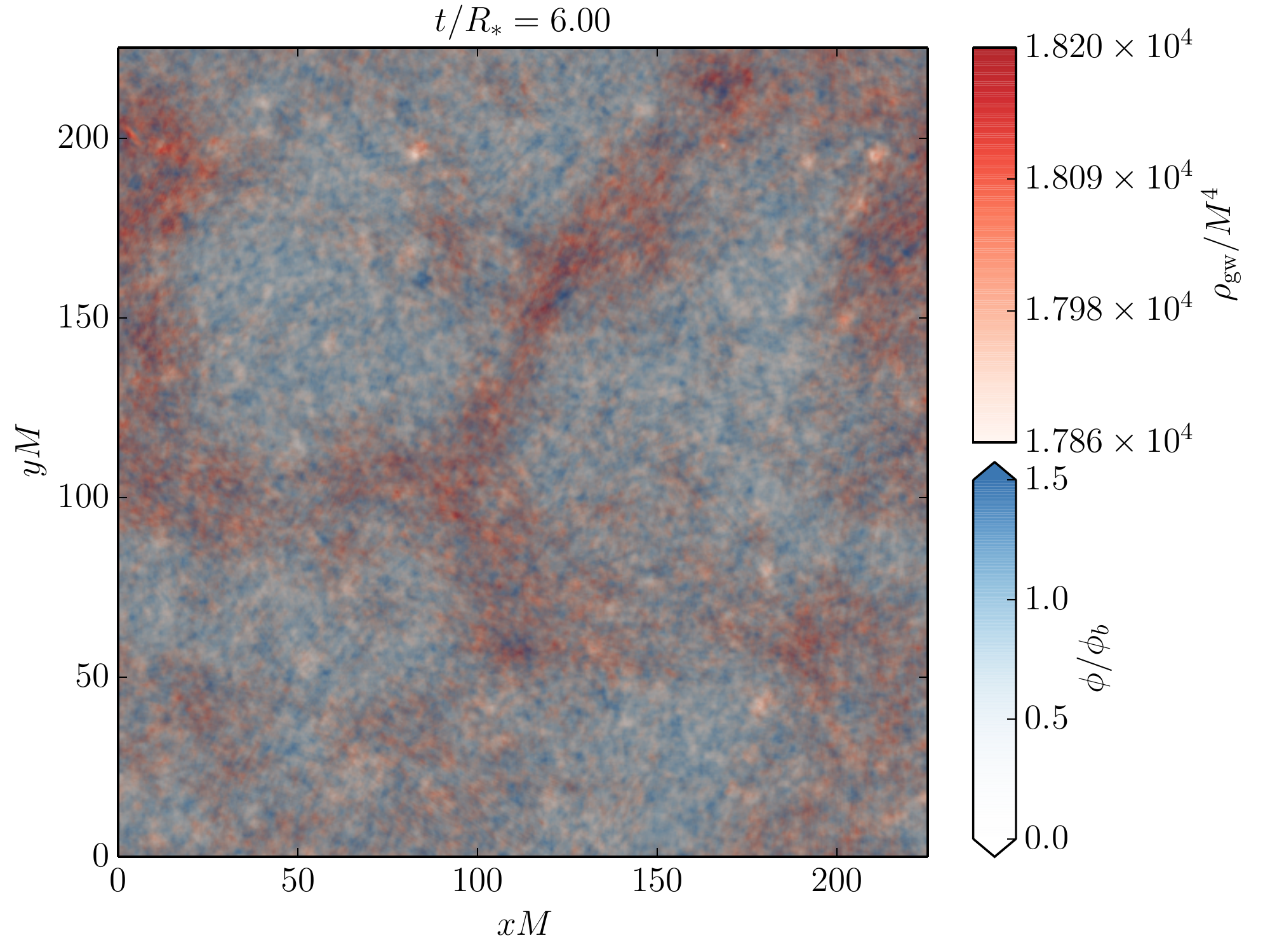}}}
\caption{Slices through a simultaneous nucleation simulation with
  parameters $\Rc M=7.15$, $\Nb=64$ and $\Rstar M=56.32$ showing the 
  expansion (a), collision (b), and oscillatory (c and d) phase of the
  scalar field.  The scalar field value is shown in blue, and the
  gravitational wave energy density is shown in red. Note that the
  range of the colourbar for the gravitational wave energy density
  changes for each plot.  During the oscillatory phase the
  gravitational wave energy density becomes very uniform and the
  ``hotspots'' are deviations on the sub percent level. The full set of parameters for this run is shown in Table~\ref{table:sim}. A movie based on this simulation is included in the supplemental material.  }
\label{fig:ScalarSlices}
\end{figure*}

There are a number of length scales within our simulation. The largest physical length scale within our system is the average separation between bubbles $\Rstar $. This is simply given by 
\begin{equation}
\Rstar =\left(\frac{\Vol}{\Nb}\right)^{1/3}\text.
\end{equation} 
Much smaller than this length scale is the radius of the critical bubble $\Rc$, and the critical bubble wall width $l_0$. The critical bubble wall width is associated with the scalar field mass in the broken phase. Smaller still is the length scale of the Lorentz contracted bubble walls. We define $\gmStar=\Rstar /2\Rc $ which is the expected Lorentz factor for a bubble with diameter $\Rstar $, and then define $l_*=l_0/\gmStar$ as the width of the Lorentz contracted bubble wall with a diameter of $\Rstar $. It is crucial that we have a good resolution of the bubble walls up until they collide, and as such we need our lattice spacing $\Delta x \ll l_*$. Note that by obtaining different values of $\Rc $ we can vary $\Rstar$ while keeping $\gmStar$ the same. 

In most vacuum phase transitions we expect bubbles to expand to many times the size of the critical bubble, and therefore up to very high Lorentz factors. We also would like to have many bubbles within our simulation box to obtain an accurate ensemble. Hence, we need sufficiently large lattices to separate the scales
\begin{equation}
\Delta x\ll l_* \ll l_0\, \lesssim\, \Rc  \ll \Rstar  \ll L\Delta x \text.
\end{equation}
It is not possible to perform a simulation in which we achieve a realistic value for $\gmStar$ and a correct separation of scales. Instead, we perform multiple simulations with increasing values of $\gmStar$ to attempt to identify a trend as $\gmStar\rightarrow \infty$. 

For simulations with a given nucleation rate, we typically expect the first bubble nucleated to grow to a larger size than bubbles nucleated later, and so the wall of the first bubble when it collides will have $\gamma$ greater than $\gmStar $. Its bubble wall at collision will therefore be thinner than $l_*$. This effect is particularly pronounced for simulations with an exponential nucleation rate where the first bubble nucleated often grows to be many times larger than the subsequent bubbles at collision time. For a simultaneous nucleation run the diameters of the bubbles will be more closely distributed around $\Rstar $, 
and so the thinnest wall at collision will be much closer to $l_*$. 

For an exponential nucleation rate simulation we need a much finer lattice spacing in comparison to a simultaneous nucleation simulation with the same $l_*$. In practice, reducing the lattice spacing is too expensive and for large volumes 
we become unable to trust our results due to bad energy conservation. 

\section{Results: scalar field}\label{sec:ResultsScalar}

As described in the previous section, 
the scalar field evolution can roughly be split into three stages, expansion, collision, and oscillation. Slices through a simultaneous nucleation simulation volume are shown in Fig.~\ref{fig:ScalarSlices}. During collision, we see many regions in which the scalar field is rebounding into the symmetric phase as described in Section \ref{sec:CollOsc}. During the oscillation phase the scalar field becomes more homogeneous on large scales while the scalar field continues to oscillate on small wavelengths. This persists 
as long as the simulations run,
for times that are many multiples of $\Rstar$.

In order to test for lattice effects, we study single bubbles, whose Lorentz factors should be related to their radius through Eq.  (\ref{eq:gamma_bubble_growth}).
We find the volume 
by counting the number of lattice sites with $\phi>\phiAtMin$, and then
from this we are able to deduce the bubble radius $R$ and the Lorentz factor factor of the wall $\ga$. 

\begin{figure}
\centering 
\includegraphics[width=0.48\textwidth]{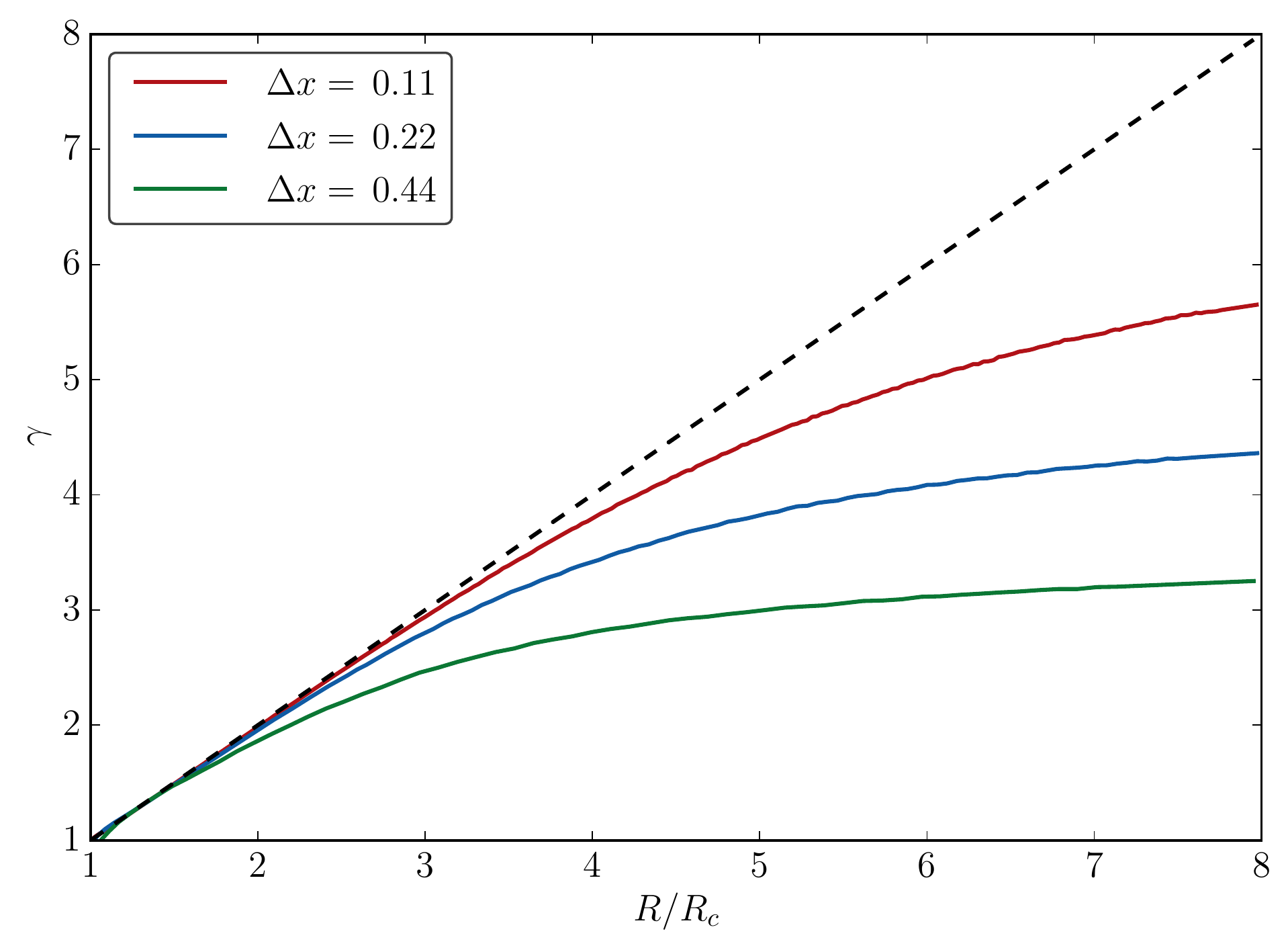}
\caption{The Lorentz factor $\gamma$ of the bubble wall for different values lattice spacings plotted against the radius of the bubble in units of the critical radius. This is for a bubble with $\Rc M=7.15$. The dashed black line shows $\gamma=R/\Rc$.  
}\label{fig:MultiBubGamR}
\end{figure}

We plot $\gamma$ against $R$ 
in Fig.~\ref{fig:MultiBubGamR}. 
The lattice effects are easy to see, 
as  $\gamma$ is highly sensitive to small changes in velocity when $\vw\rightarrow 1$. 
The bubble wall is stopped from contracting beyond a width which is representable on the lattice, 
and the bubble wall is unable to increase its velocity. 
The energy that is lost is transferred to small wavelength oscillations that follow behind the bubble wall. 
This effect has been seen previously in accelerating kinks on a lattice \cite{PhysRevB.28.6873,Peyrard:1985uf}.

If the deviation of $\gamma$ from its theoretical value becomes sufficiently large then this can 
be associated with loss of energy conservation. 

\begin{figure}
\centering 
\includegraphics[width=0.48\textwidth]{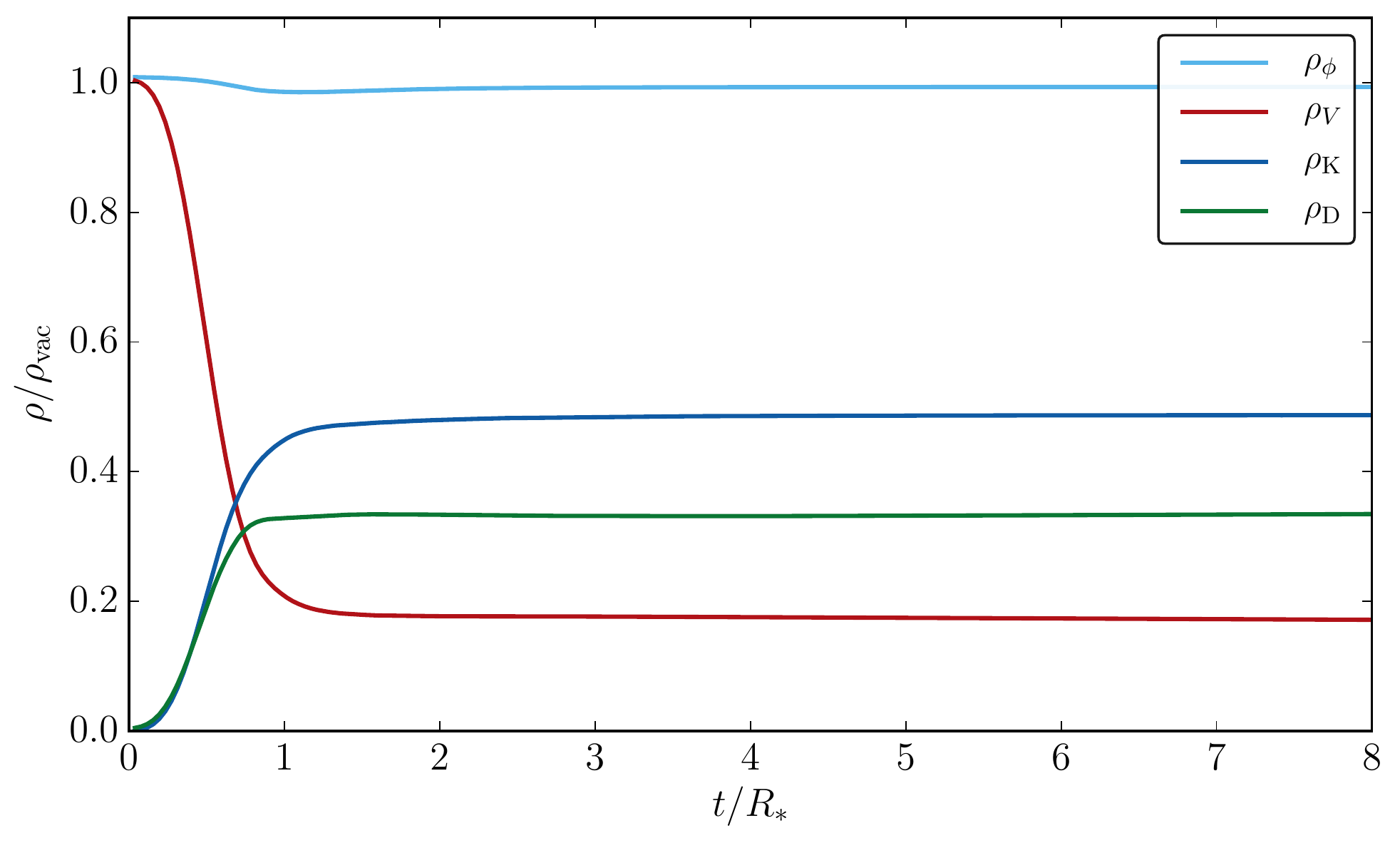}
\caption{Energy densities in the scalar field over time for a
  simultaneous nucleation run with $\Rc M=7.15$, $\Rstar M=56.3$ and
  $\Nb=4096$. The full set of parameters for this run is shown in Table~\ref{table:sim}.}
\label{fig:EnergyDens}
\end{figure}

We plot the energy densities over time for a simultaneous nucleation phase transition in Fig.~\ref{fig:EnergyDens}. As the bubbles expand the potential energy drops steeply and the kinetic and gradient energies increase. Initially the gradient energy and kinetic energy are roughly equal but when the bubbles begin to collide the kinetic energy becomes larger than the gradient energy. Shortly after the phase transition enters the oscillation stage, with $\rho_V\neq 0$. 

\begin{figure}
\centering 
\includegraphics[width=0.48\textwidth]{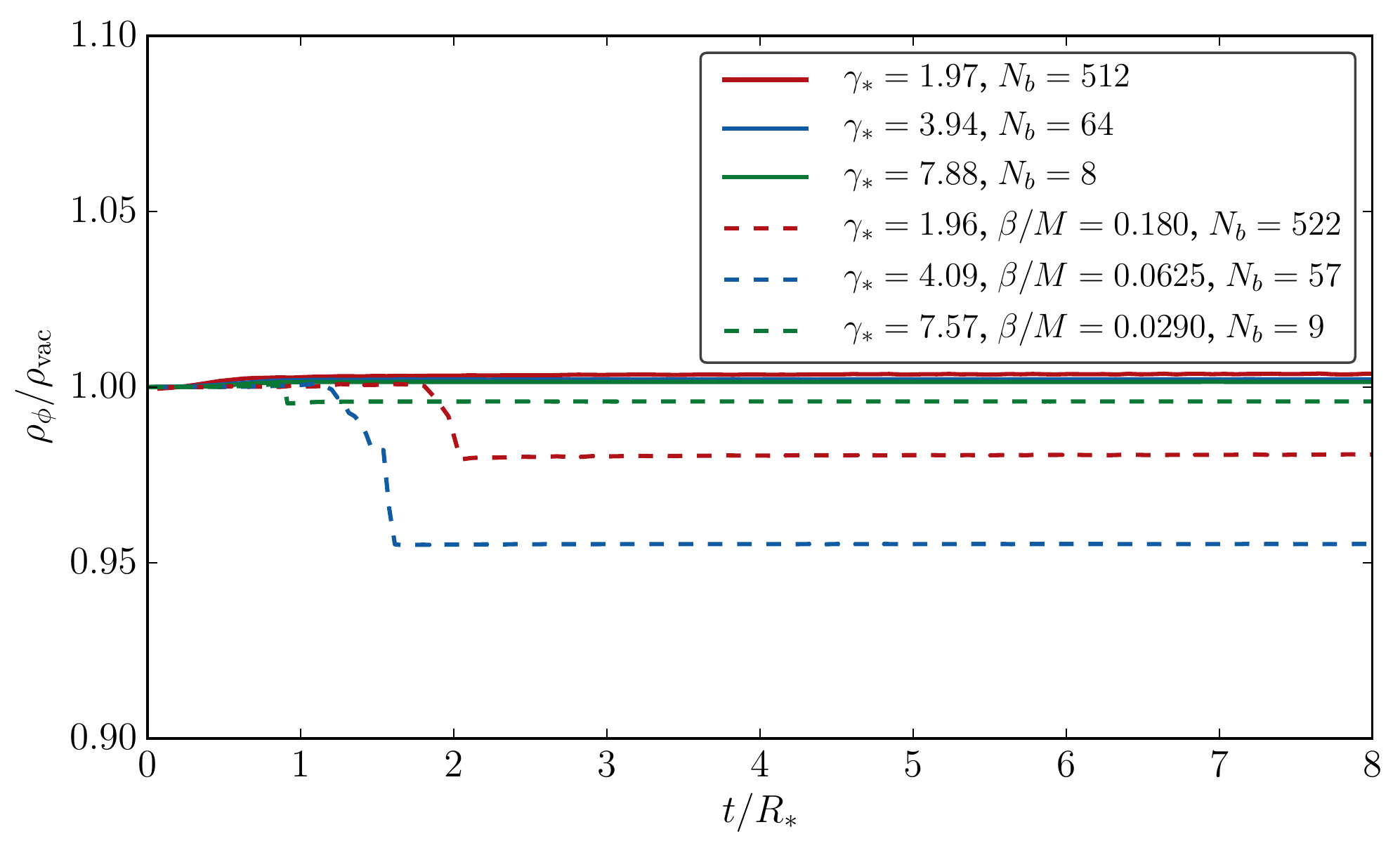}
\caption{Energy conservation for several simulations of the same physical volume. Runs with exponential nucleation are plotted with dashed lines, and simultaneous nucleation runs are shown with solid lines. 
See Tables \ref{table:sim} and \ref{table:exp} for the full set of parameters of each run.
} \label{fig:EnergyCons}
\end{figure}

The energy conservation for a series of simulations with $L\Delta x
=225.28$ is given in Fig.~\ref{fig:EnergyCons}. We can see that energy
conservation is substantially better in the simultaneous nucleations
in comparison to the exponential nucleation runs. This is what we
expected due to the biggest bubble/thinnest wall effect mention in
section \ref{sec:Methods}. These are the largest volume simulation
runs for exponential nucleation, and so have the worst energy
violation of all simulations performed. Even in the worst case, energy
conservation violation is still kept to $\lsim 5\%$.

To monitor energy conservation in our multi-bubble simulations, 
we define a new parameter $\gmStarLat$ which is the 
numeric value found for $\gamma$ on the lattice when the bubble radius is 
$R=R_*$. This new parameter is listed for all simulation runs in their respective tables.

\begin{figure*}
\centering 
\subfigure[\ Simultaneous]{\includegraphics[width=0.485\textwidth]{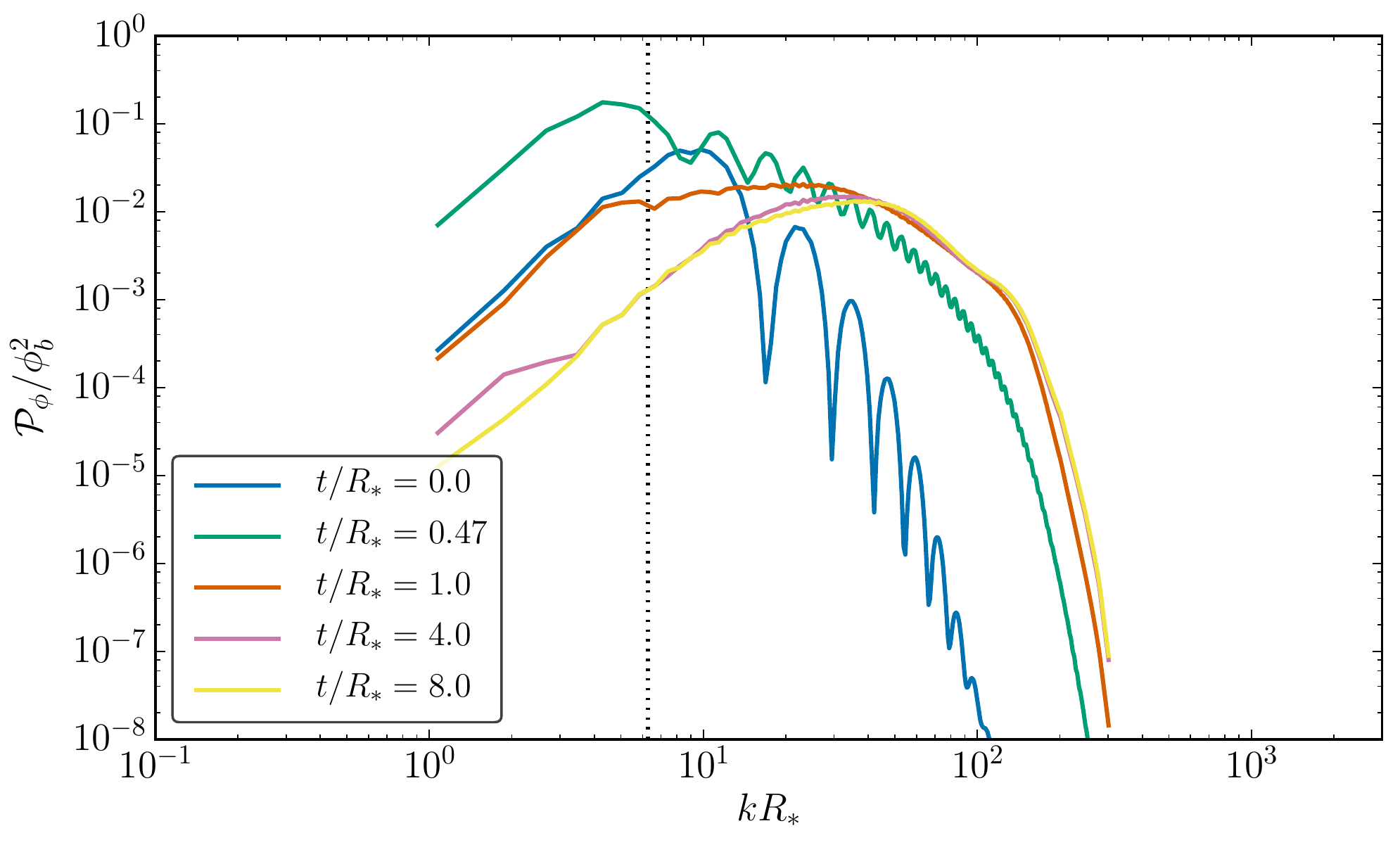}}
\hfill
\subfigure[\ Exponential]{\includegraphics[width=0.485\textwidth]{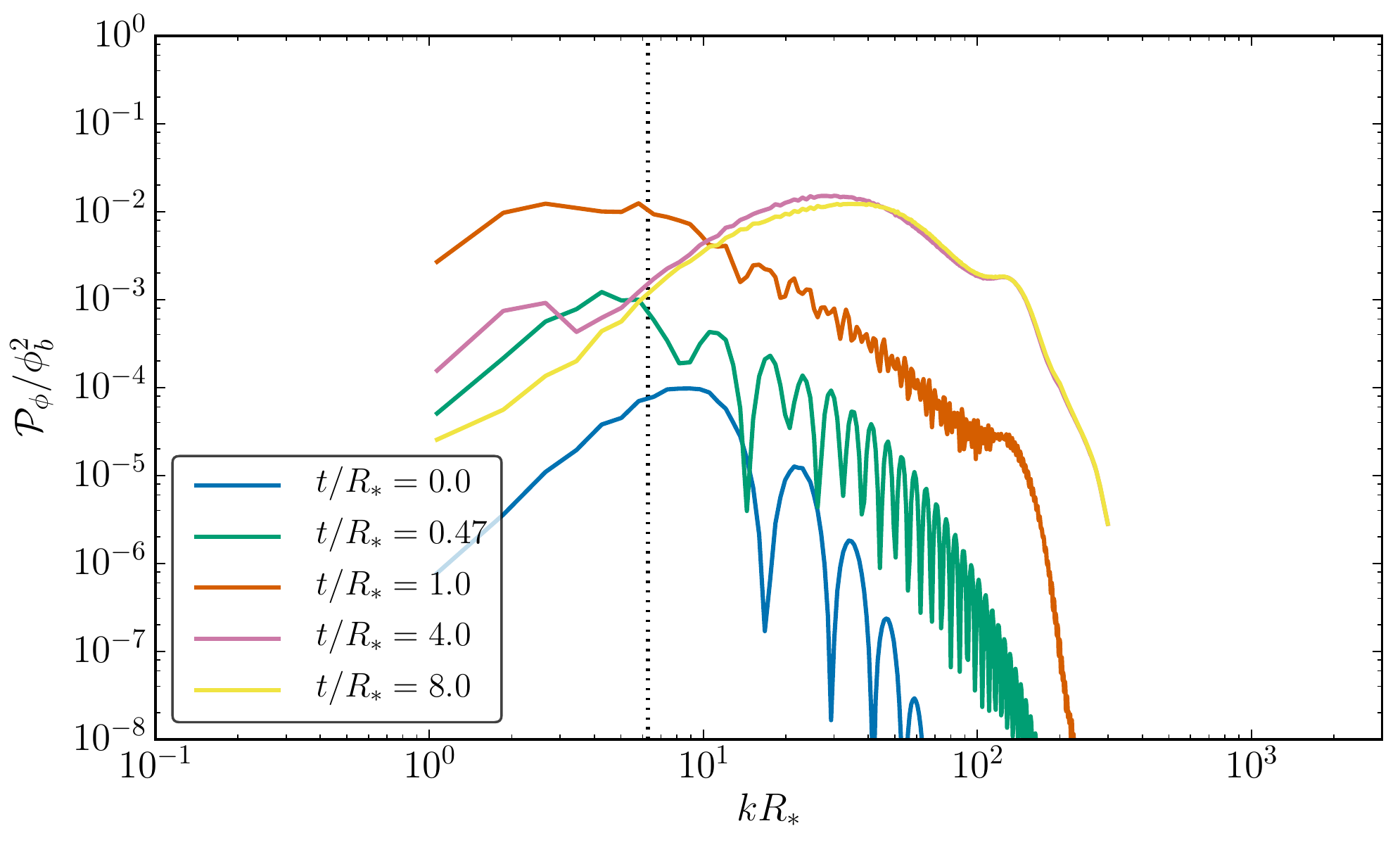}}
\caption{Scalar power spectra for simultaneous (left) and exponential (right) nucleation runs. Both simulations have $\Rc M=7.15$. The left plot has $\Nb=512$ and $\gmStar=1.97$, while the right plot has $\Nb=522$ and $\gmStar=1.96$. The full set of parameters of each run can be extracted from Tables \ref{table:sim} and \ref{table:exp}.
The initial configuration of the scalar field is seen at $t/\Rstar =0$. The bubble expansion phase is seen for $t/\Rstar =0.47$. The spectrum during bubble collision is seen at $t/\Rstar =1.0$. The late time power spectrum after bubbles have collided is then shown at $t/\Rstar =4.0$ and $t/\Rstar =8.0$. The vertical black dotted line denotes $k=2\pi/\Rstar$.} \label{fig:ScalarSpecEvol}
\end{figure*}

The power spectrum of the scalar field $\mathcal{P}_\phi$ can inform us about the 
length scales
of the shear stresses sourcing gravitational waves. We plot $\mathcal{P}_\phi$ during the expansion, collision and oscillation phases for both a simultaneous and exponential nucleation run in Fig.~\ref{fig:ScalarSpecEvol}. During expansion and collision $\mathcal{P}_\phi$ is peaked around $\Rstar$. During the oscillation phase $\mathcal{P}_\phi$ shifts so that its maximum is at a higher wavelength, closer to the scale associated with $l_0$. This can be understood as the structure of bubbles disappearing and being replaced with oscillating features around the mass scale of the scalar field.

The main differences in $\mathcal{P}_\phi$ between the simultaneous and exponential nucleation runs are during the expansion and collision phases. Identical bubbles are all spawned at the start of the simultaneous nucleation run, and so $\mathcal{P}_\phi$ has a larger magnitude at $t/\Rstar=0.0$, 
and shows the characteristic ``ringing'' of the single-bubble power spectrum.
These bubbles then expand in a uniform way, their geometries differentiating from each other only upon collision with another bubble. 
Comparatively, as more bubbles are spawned during the exponential nucleation run $\mathcal{P}_\phi$ 
becomes smoother on large scales and noisier on small scales
as bubbles of varying sizes appear. 
The collision phase also lasts longer,  and 
the distribution of bubbles is not as homogeneous as in a simultaneous nucleation run.

\begin{figure}
\centering 
\includegraphics[width=0.48\textwidth]{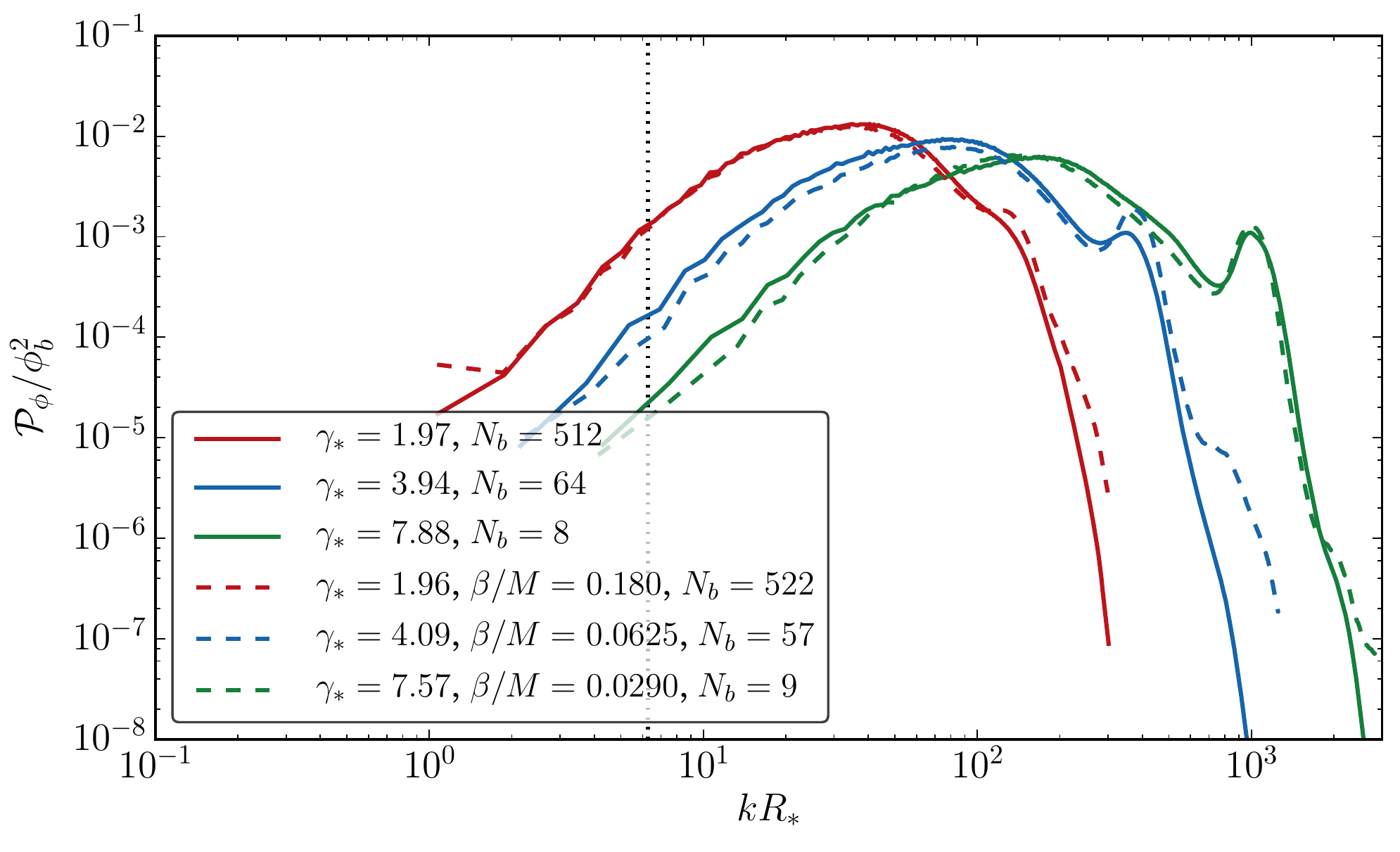}
\caption{Comparison of the late time scalar power spectrum at $t/\Rstar=8.0$ for both simultaneous (solid lines) and exponential (dashed lines) nucleation runs. All runs use bubbles with $\Rc M=7.15$.
See Tables \ref{table:sim} and \ref{table:exp} for the full set of parameters of each run.
}\label{fig:ScalarSpecBoth}
\end{figure}

If we plot the late time scalar field power spectrum together on the same graph we can clearly see that during the oscillation phase the simultaneous and exponential nucleation runs settle into 
similar 
states. We do this for several values of $\gmStar$ in Fig.~\ref{fig:ScalarSpecBoth}. In all cases, the scalar power spectra settle into similar states apart from lattice effects in the UV.

\section{Results: gravitational waves}\label{sec:ResultsGrav}

\subsection{Simultaneous nucleation}

In Fig.~\ref{fig:ScalarSlices}, we also show in shades of red 
the gravitational wave energy density $\rGW(\bx,t)$ sourced by the scalar field. 
To obtain $\rGW$ in real space we first perform
the Fourier transform of $\dot{u}_{ij}$, then project this to obtain
$\dot{h}^{TT}_{ij}$ in $k$-space. Finally we perform the inverse Fourier
transform to find $\dot{h}^{TT}_{ij}$ in real space. From this we then
calculate $\rGW(\bx,t)$ using Eq.~(\ref{e:GWEnDen}).

We can clearly see from
Fig.~\ref{fig:ScalarSlices} that during the collision phase hotspots
in $\rGW$ are located in regions where bubbles are colliding. These
are the locations where the spherical symmetry of the expanding
bubbles is broken. During the oscillation phase the gravitational wave
energy density becomes largely homogeneous with fluctuations on the
percent level, though gravitational waves continue to be sourced. 

\begin{figure}
\includegraphics[width=0.48\textwidth]{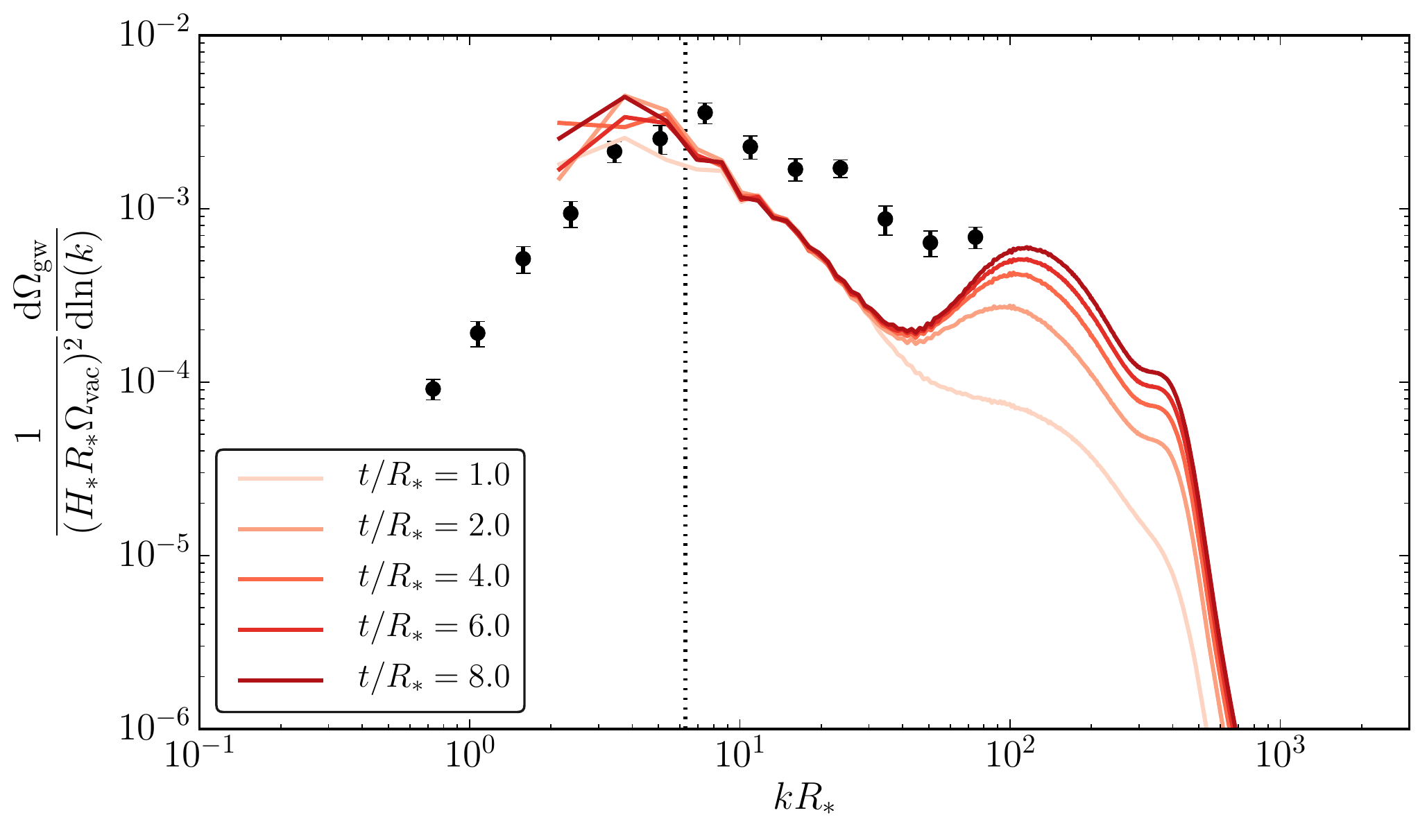}
\caption{The gravitational wave power spectrum for the simultaneous
  nucleation run with $\Rc M=7.15$, $N_b=64$ and $\gmStar=3.94$ listed in
  Table~\ref{table:sim}. The vertical black dotted line marks where
  $k=2\pi/\Rstar$. The black data points are the results for running a simulation with the envelope approximation with the same bubble locations and nucleation times. 
} \label{fig:OmgSpecSimFit}
\end{figure}

In Fig.~\ref{fig:OmgSpecSimFit}, we plot the 
gravitational wave power spectrum 
at several times 
over the duration of a simultaneous
nucleation simulation with $\gmStar\simeq 4$. As the bubbles begin to
collide we begin to see a peak in the spectrum emerging near
$k=2\pi/\Rstar$, with 
a power law 
fall-off towards the UV. For this simulation we do not have
a sufficient separation between $R_*$ and $L\De x$ 
to estimate the power law towards
the IR. As the collision phase completes this peak and the power law
towards the IR persists, but a second peak associated with a much
smaller length scale continues to grow. This second peak is due to
gravitational waves sourced from oscillations in the scalar field 
with wavelengths close
to the inverse mass of the scalar field. 

In the same figure, we also plot the results of a 
numerical calculation in the envelope approximation, as
detailed in \cite{Weir:2016}, using the same bubble nucleation locations.
The peak power  
in our simulation
is closely reproduced by the envelope calculation, although the 
envelope calculation predicts that the peak is at higher frequency.
The power law towards the UV is somewhat steeper in the numerical simulations 
than in the envelope calculation.

\begin{figure}
\centering 
\includegraphics[width=0.48\textwidth]{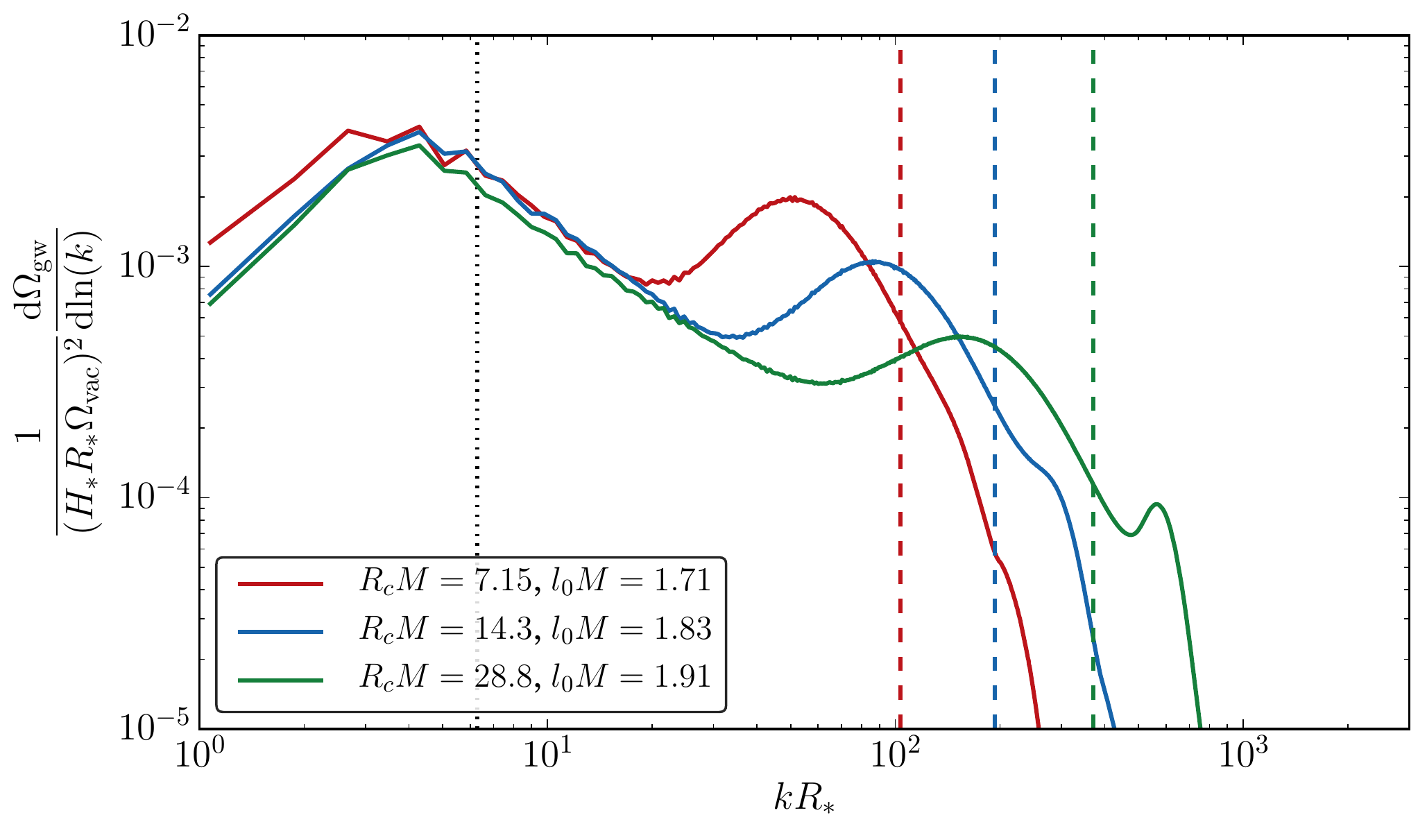}
\caption{Gravitational wave power spectrum for several runs with
  different critical radius $\Rc $ and $\Rstar $. For each simulation
  the power spectra have been averaged over the interval $2.5\leq
  t/\Rstar \leq 8.0$. All simulations shown have the same number of
  bubbles $N_b=512$ and $\gmStar\simeq 2 $, 
  with the full set of parameters listed in Table \ref{table:sim}.
  We also plot the
  length scale associated with $\Rstar $ as the vertical black dotted line, and the corresponding length scale for the initial wall width $l_0$ for each simulation as coloured dashed lines. }\label{fig:OmgSpDeltas}
\end{figure}

We can show that the frequency of peak power is 
associated with the length scale of $\Rstar$, whereas the 
bump in the UV is associated with the length scale $l_0$. In
Fig.~\ref{fig:OmgSpDeltas}, we plot the power spectra for three runs
with different values of $\Rc$ and $\Rstar$ but the same $\gmStar$. 
It can be seen that the power peaks at $kR_* \simeq 3$, with a secondary 
peak at $k l_0 \simeq 3$.

In a realistic transition, the separation between $\Rstar$ and $l_0$ will
be many orders of magnitude, and we would expect 
the UV peak will be greatly
suppressed due to the fall-off of the power spectrum with increasing
$k$. We will estimate how large it can grow below.

Note that 
the power spectrum fluctuates due to the oscillations in the individual Fourier modes. 
In order to minimise this effect, in 
some plots we average over power spectra produced during an interval spanning several
$t/\Rstar$. On these occasions the details are given in the caption of
the figure.

\begin{figure}
\centering 
\includegraphics[width=0.48\textwidth]{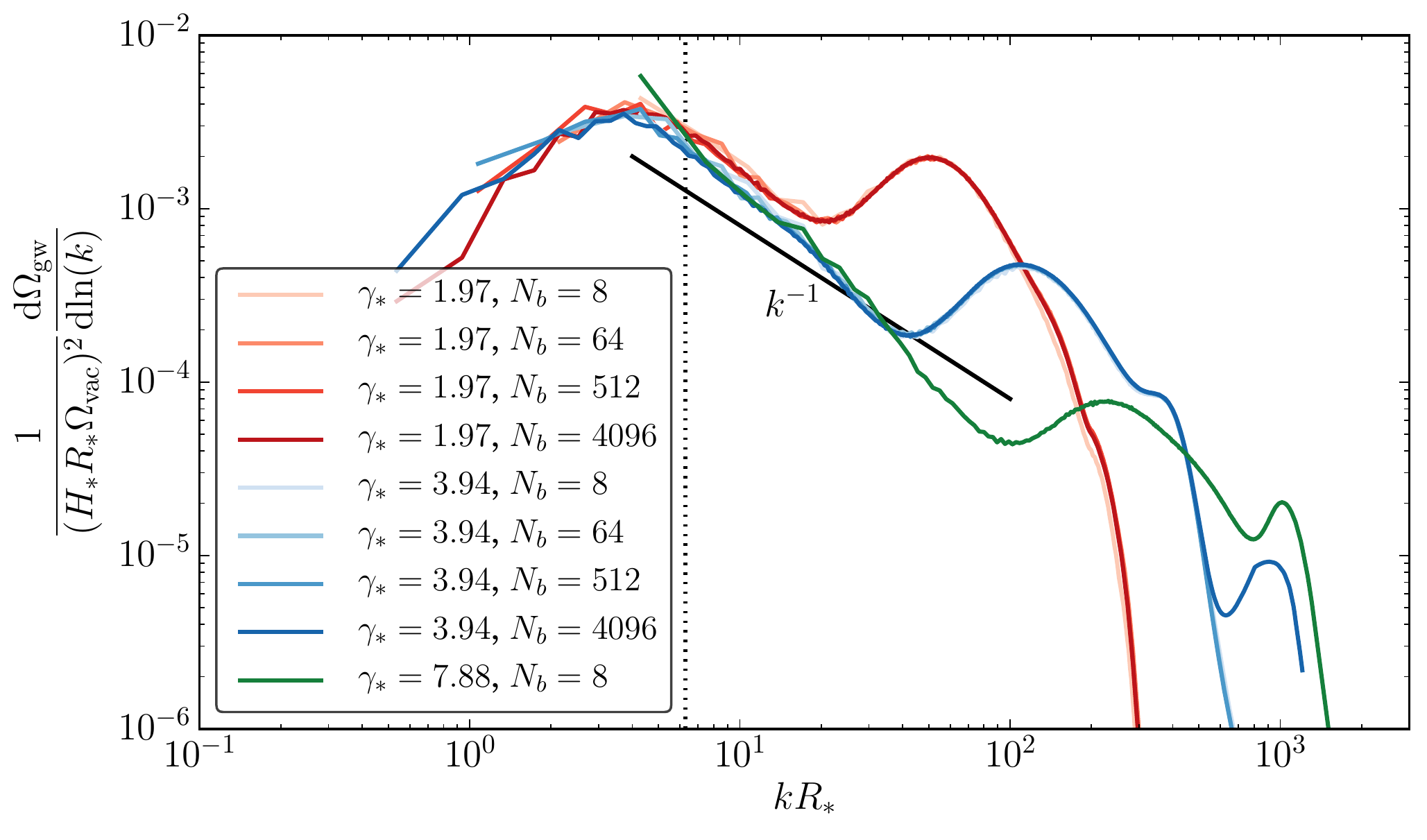}
\caption{Gravitational wave power spectrum for all simultaneous bubble
  runs with $\Rc M=7.15$. The parameters for these runs are given in
  Table~\ref{table:sim}. For each simulation the power spectra have
  been averaged over the interval $2.5\leq t/\Rstar \leq 8.0$. The solid black line shows a power law of $k^{-1}$. We plot
  as a vertical black dotted line the wave number $k = 2\pi/\Rstar$.
  See Table \ref{table:sim} for the full set of parameters of each run.
  }\label{fig:OmgSpMultSim}
\end{figure}

We show the runs with $\Rc M=7.15$ from Table~\ref{table:sim} in Fig.~\ref{fig:OmgSpMultSim}. By increasing $\Nb$ while keeping $\Rstar$ the same we are able to see further into the IR for a given $\gmStar$. Apart from this increasing, $\Nb$ does not have a significant effect on the shape of the power spectrum, implying that $\Nb=8$ is sufficient to measure the slope of the power law towards the UV. Increasing $\gmStar$ does not change the location or amplitude of the IR peak in respect to $\Rstar$. While the slope of the power spectrum towards the IR is in agreement with $k^{-1}$ for $\gmStar\simeq 2$, it appears steeper for $\gmStar \simeq 4$ and $\gmStar \simeq 8$. Between $\gmStar \simeq 4$ and $\gmStar \simeq 8$ the slope appears consistent.

\subsection{Exponential and constant nucleation}

\begin{figure}
\includegraphics[width=0.48\textwidth]{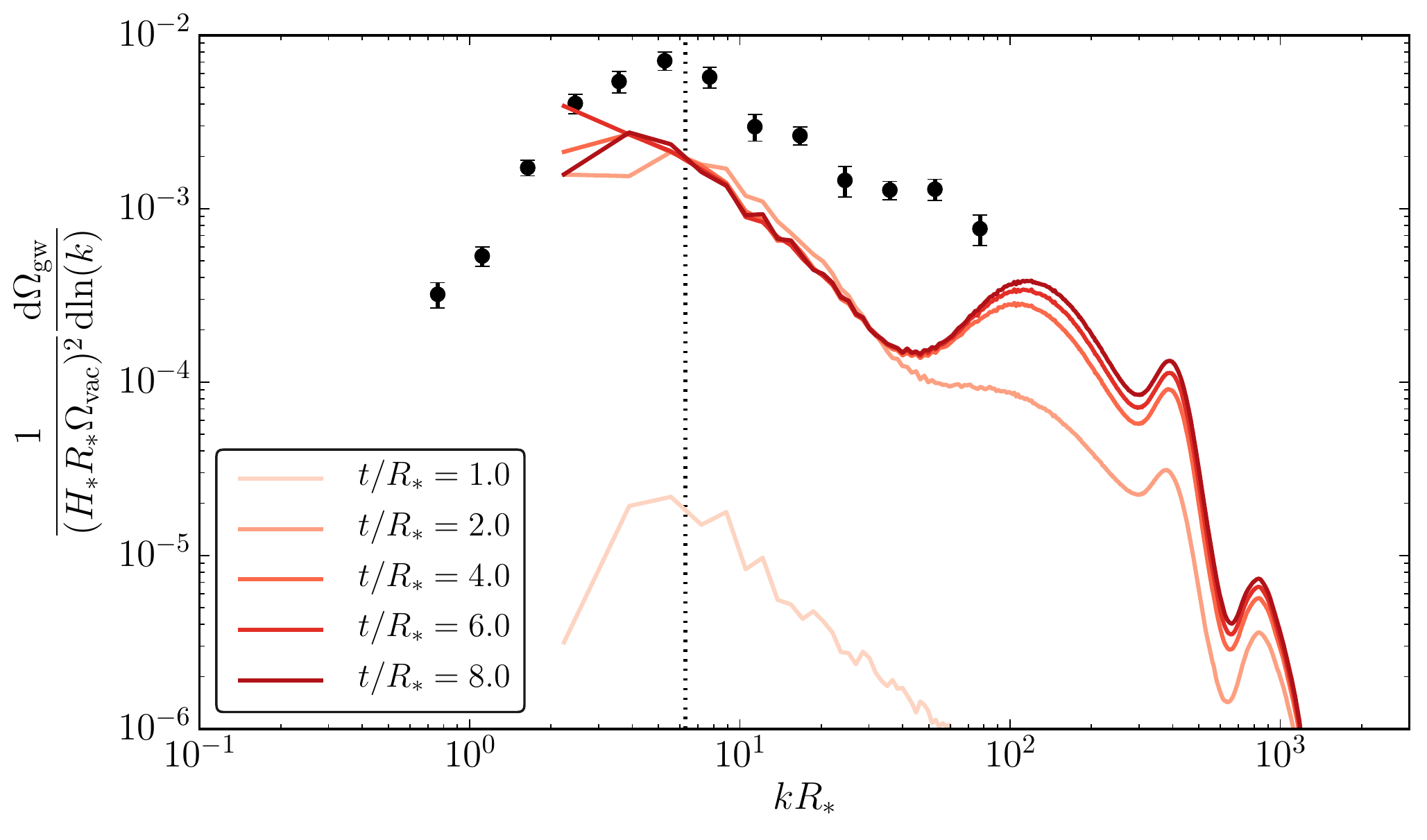}
\caption{Gravitational wave power spectrum for the exponential
  nucleation run with $\Rc M=7.15$, $N_b=57$ and $\gmStar=4.09$ listed in
  Table~\ref{table:sim}. The vertical black dotted line marks where
  $k=2\pi/\Rstar$. The black data points are the results for running a simulation with the envelope approximation with the same bubble locations and nucleation times.   }\label{fig:OmgSpecExpFit}
\end{figure}

In Fig.~\ref{fig:OmgSpecExpFit}, we show the evolution of the power
spectra for an exponential nucleation run with
$\gamma_*\simeq4$. Similar to in Fig.~\ref{fig:OmgSpecSimFit} we plot the results of a simulation using the envelope approximation as
detailed in \cite{Weir:2016} using the same bubble nucleation
locations and times. 

 For the exponential nucleation run we
see that the envelope simulation gives an
overestimate of peak amplitude, but is still within an order of
magnitude. From the full scalar field simulation we obtain a similar peak amplitude as in the
simultaneous nucleation run shown in
Fig.~\ref{fig:OmgSpecSimFit}. This indicates that the scaling of
gravitational wave production for our simulations is governed by
$\Rstar$ rather than $\beta$. Once again the peak location is shifted slightly
into the IR in comparison to the envelope simulation.

\begin{figure}
\centering 
\includegraphics[width=0.48\textwidth]{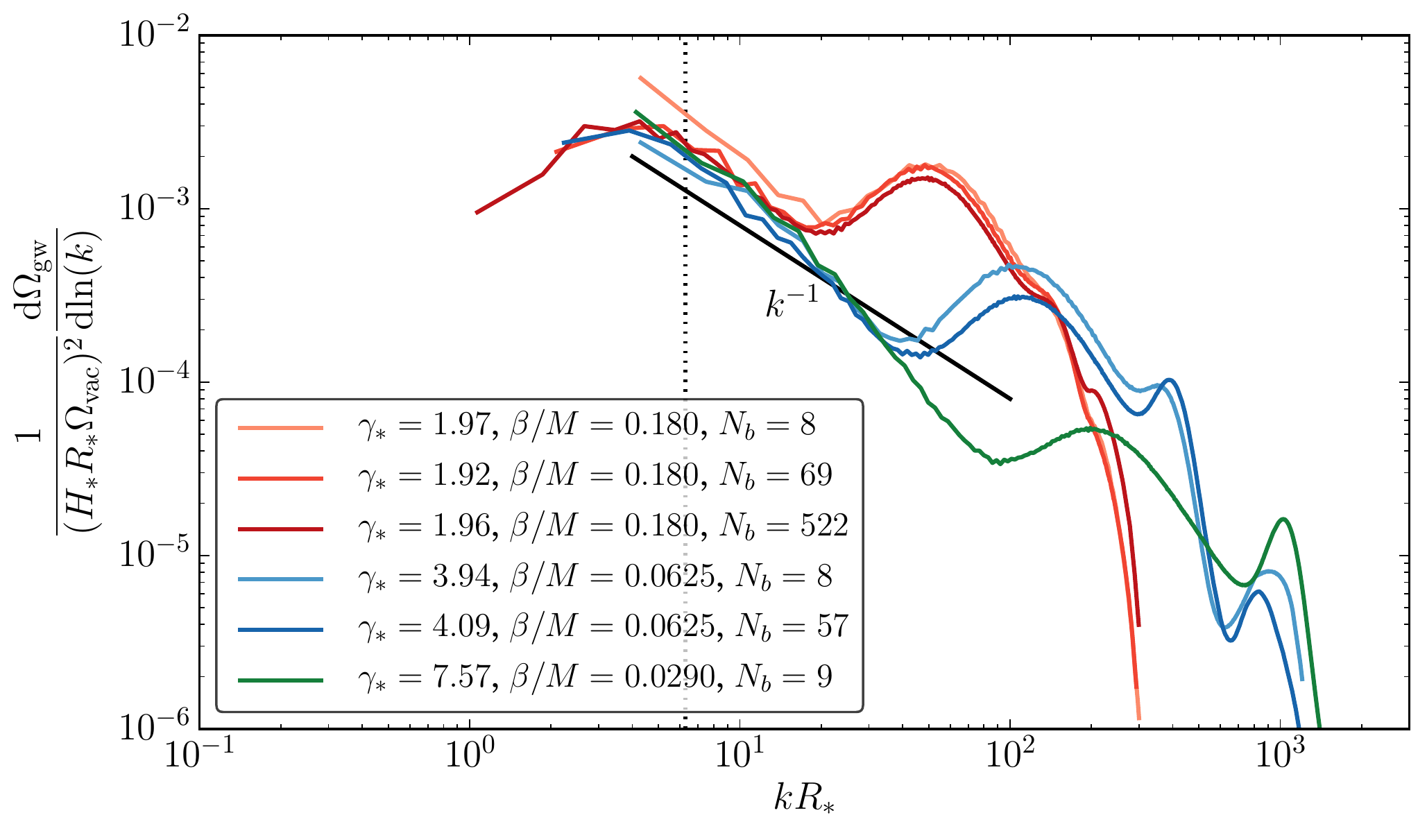}
\caption{Gravitational wave power spectrum for all exponential nucleation runs with $\Rc M=7.15$. The parameters for these runs are given in Table~\ref{table:exp}.  For each simulation the power spectra have been averaged over the interval $2.5\leq t/\Rstar \leq 8.0$. The solid black line shows a power law of $k^{-1}$.
Also plotted as a vertical black dotted line is $k\Rstar = 2\pi $.
}\label{fig:OmgSpecMultiExpAvg}

\end{figure}

The power spectra for all exponential simulation runs are shown in
Fig.~\ref{fig:OmgSpecMultiExpAvg}. For similar $\gmStar$, we see convergence to the resulting slope of the power spectra for even small numbers of bubbles, implying that even $N_b=8$ creates a satisfactory ensemble.

All simulations seem to be consistent
regarding the location and height of the peak in the IR and there is
even agreement with the simultaneous nucleation runs. The slope of the
power spectrum towards the IR is steeper than $k^{-1}$ for
$\gmStar\simeq 4$ and $\gmStar \simeq 8$, and appears consistent
between them. 

The two constant nucleation runs listed in Table~\ref{table:con} are
found to produce power spectra that are consistent with the
simultaneous and exponential nucleation runs. We plot the power
spectra for the constant nucleation runs along with the other
$\gmStar\simeq 4$ runs in Fig.~\ref{fig:OmgSpecMultiGam4Avg}.

\subsection{Late time power spectrum}
We are able to see the shape of the power spectrum generated during the oscillation phase by setting $u_{ij}=0$ after the collision phase has completed. We chose a time $t/\Rstar=2$ to set $u_{ij}=0$. However for some simulations there appears to have been regions in which bubbles were still colliding at this time, and so a later time should have been chosen. For these simulations there is an uptick in the power spectrum in the IR, which can contribute significantly to the energy density. As the IR bins consist of only a few modes, there can be  
large oscillations in $\OmGW$.

\begin{figure}
\includegraphics[width=0.48\textwidth]{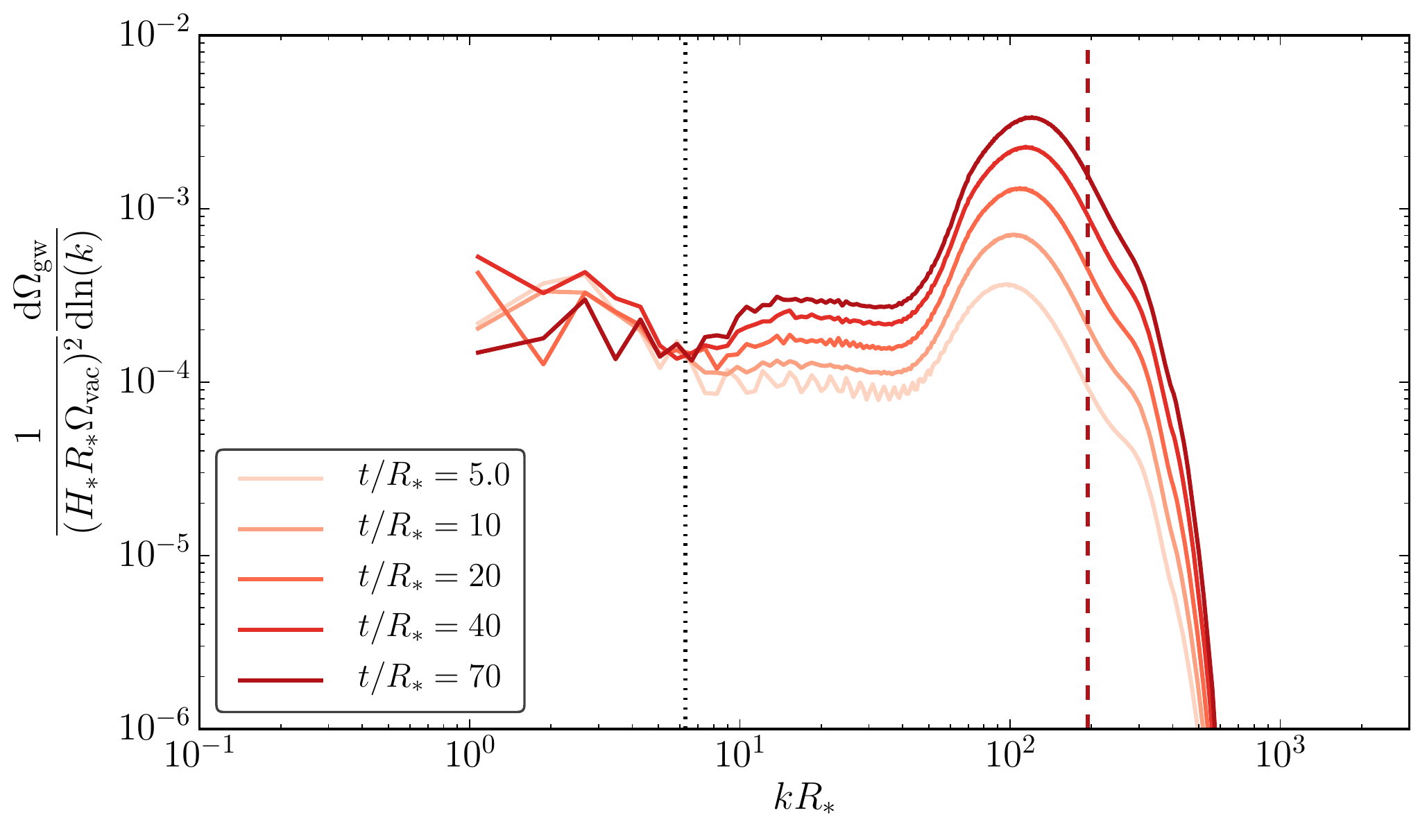}
\caption{Late time power spectrum from gravitational waves generated
  from the oscillation stage. Metric perturbations are only turned on
  after $t/\Rstar=2$. This is for the run with $\Rc M=14.3$, $N_b=512$
  and $\gmStar=3.94$ listed in Table~\ref{table:late-sim}, where the full set of parameters of this run are given. The vertical
  black dotted line designates where $k=2\pi/\Rstar$ and the dashed red line shows where $k=2\pi/l_0$.
  }\label{fig:OmSpLs}
\end{figure}

The evolution of the power spectrum from the oscillation phase is shown in Fig.~\ref{fig:OmSpLs}. The spectrum consists of a bump in the UV corresponding to the length scale of $l_0$ and also a plateau extending from the bump up to just before the length scale of $\Rstar$ in the IR. 
A similar shape can perhaps be discerned
in Ref.~\cite{Child:2012qg}, where the contribution to the total power spectrum 
from the oscillation phase appears to dominate. In the
  aforementioned study, the gravitational power
spectrum from collisions was 
estimated
to be between two and three orders of
magnitude smaller than that predicted by the envelope
approximation. 
The reason for this deficit is unclear. 
There was also a relatively small scale 
separation between $\Rstar$ and $l_0$, as $\gmStar$ ranges between $\gmStar\simeq2$
and $\gmStar\simeq3$. Together these may explain why the contribution
from the oscillation phase dominated that of bubble collisions.

\begin{figure}
\centering 
\includegraphics[width=0.48\textwidth]{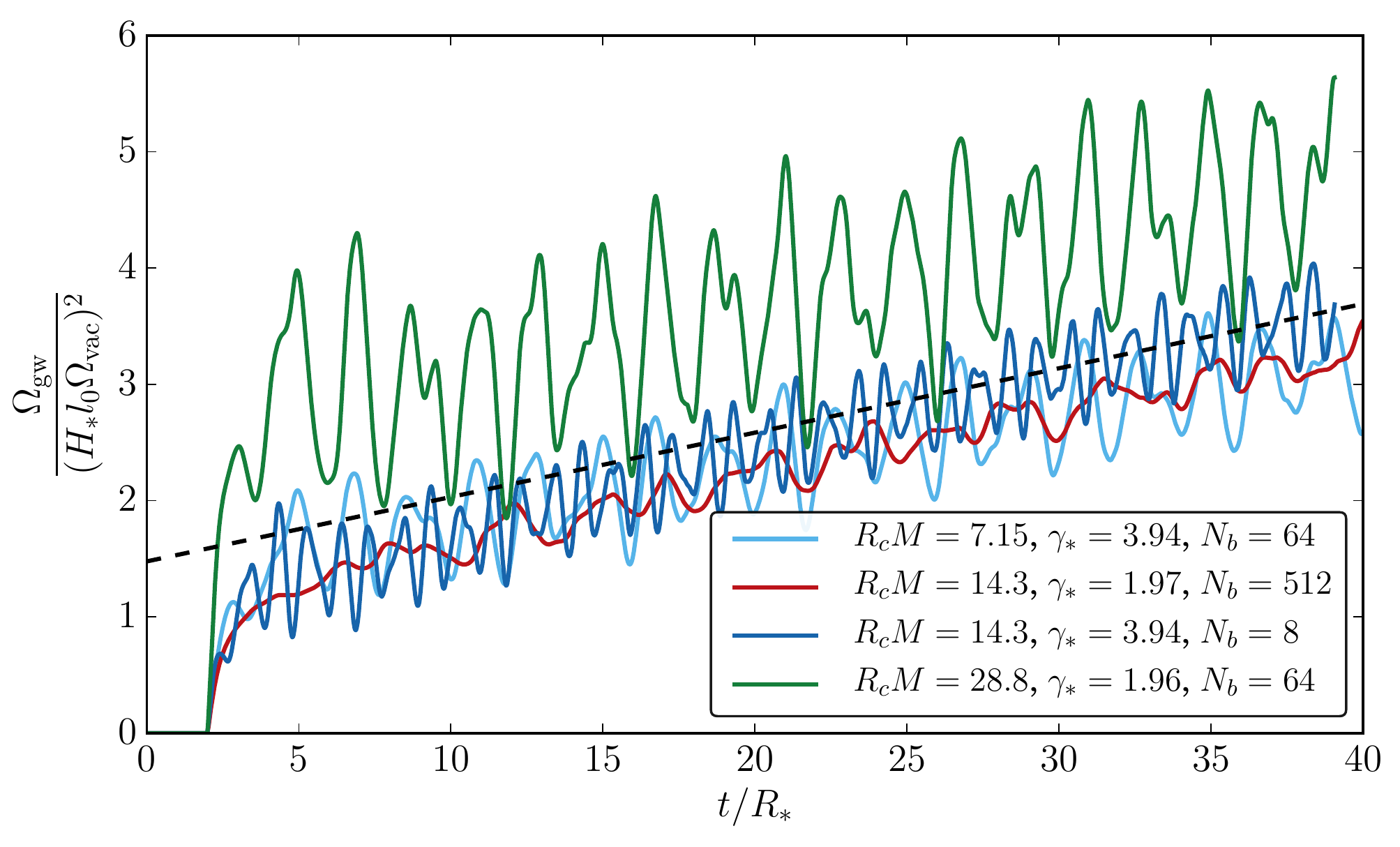}
\caption{Total $\OmGW$ from gravitational waves generated after $t/\Rstar =2$ for a series of simulations with different $\Rc $, $N_b$ and $\gmStar$, see Table~\ref{table:late-sim}. The oscillations are due to ringing in the IR of the power spectrum. 
The dashed black line is a fit for the rate of increase of
$\OmGW$, with a slope of $d\OmGW/dt=0.06 \,(\Hc l_0 \OmVac)^2/{\Rstar} $.
}
\label{fig:OmUBTot}
\end{figure}

We can see that the power spectrum continues to grow during the oscillation phase. One might 
conclude
that the contribution from the oscillation phase would eventually dominate that from the bubble collisions. 
We therefore plot $\OmGW$ for a set of simulations where the metric perturbations are turned on after $t/\Rstar=2.0$ in Fig.~\ref{fig:OmUBTot}. We are able to estimate the growth of $\OmGW$ during the oscillation phase from these simulations. We find that
\begin{equation}
\frac{d\OmGW^\text{osc}}{dt} \sim 10^{-1} \frac{(\Hc l_0 \OmVac)^2}{\Rstar}\text.
\end{equation}

The largest amount of time that $\OmGW^\text{osc}$ can grow before the
growth is cut off by expansion \cite{Hindmarsh:2015qta} is one Hubble time $\Hc^{-1}$.

From our earlier plots we can estimate the contribution to $\OmGW$ from the bubble collision phase is
\begin{equation}
\OmGW^\text{coll}\sim 10^{-3} (\Hc \Rstar \OmVac)^2\text.
\end{equation}
Therefore the ratio between these two contributions is 
\begin{align}
\frac{\OmGW^\text{coll}}{\OmGW^\text{osc}}&\sim 10^{-2} (\Rstar\Hc)^3
\frac{1}{(l_0 \Hc)^2} \text, \\
&\sim  10^{-3} \frac{\Hc^3}{n_\text{b}}
\frac{(\Mb m_\text{Pl})^2}{\rho_c} \text,
\end{align}
where 
$m_\text{pl}$ is the Planck mass. For a vacuum dominated phase
transition $\rho_c\sim \rho_\text{vac} < \frac{1}{12\lambda}\Mb^4$.
\begin{equation}
\frac{\OmGW^\text{coll}}{\OmGW^\text{osc}}\gtrsim 10^{-1}
\frac{\Hc^3}{n_\text{b}}
\left(\frac{m_\text{Pl}}{\Mb}\right)^2 \text.
\end{equation}
Providing that the mass scale of the phase transition is
sufficiently smaller than the Planck scale, the contribution from the
collision phase should dominate.

\subsection{Fitting}

\begin{figure}
\centering 
\includegraphics[width=0.48\textwidth]{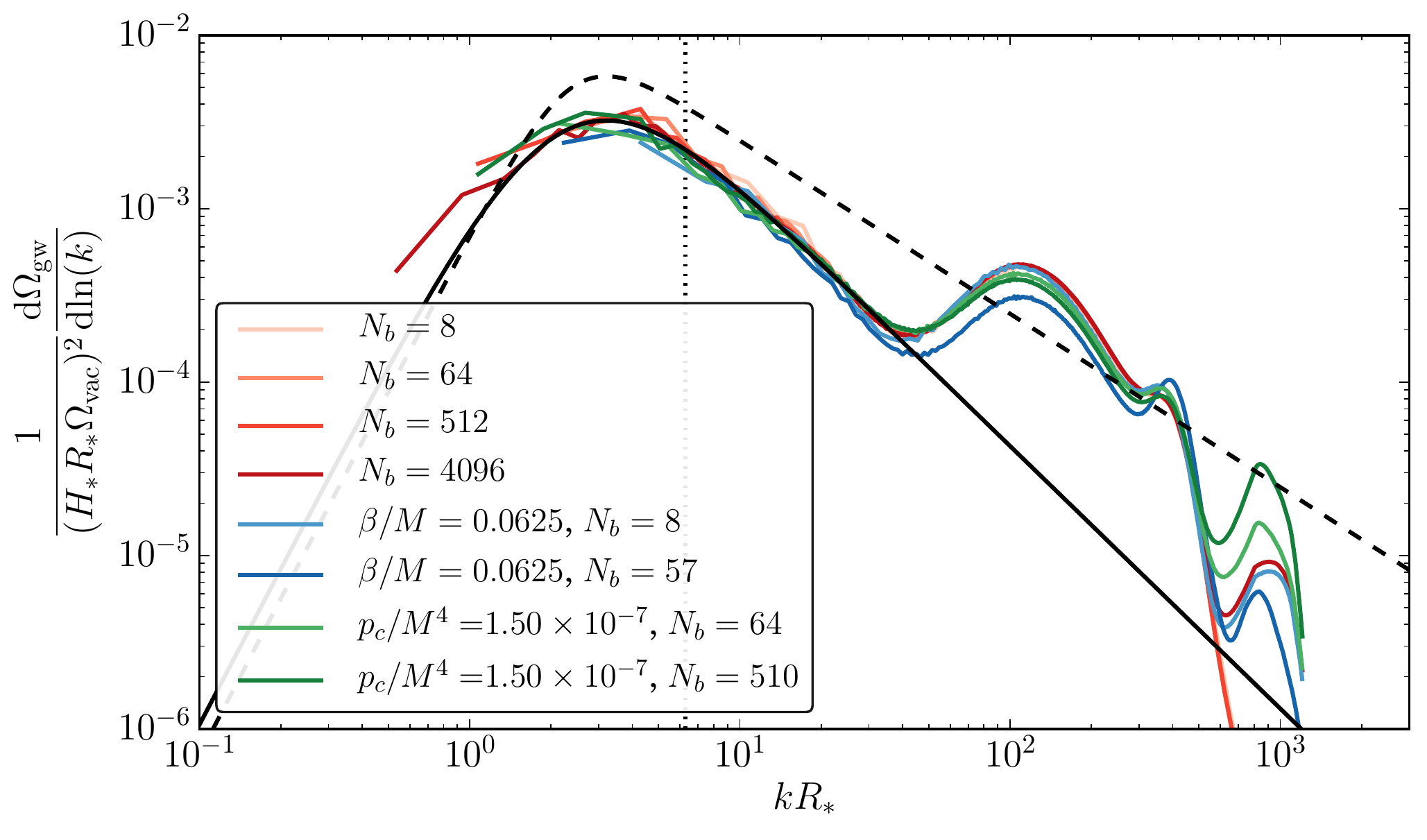}
\caption{Scaled gravitational wave power spectrum for
all simulations with $\gmStar\simeq 4$. For each simulation the power
spectra have been averaged over the interval $2.5\leq t/\Rstar \leq
8.0$. Simultaneous nucleation runs are plotted in red, exponential nucleation in blue, and constant nucleation in green.
From these simulation runs we
make a fit for the gravitational wave power spectrum from bubble
collisions given $\Rstar $, which is shown as the black solid
line. The envelope approximation fit as given in
\cite{Konstandin:2017sat} is shown as the dashed black line, where we
have used Eq.~(\ref{eq:betaGivenRstar}) to convert between $\beta$ and
$\Rstar$.}\label{fig:OmgSpecMultiGam4Avg}
\end{figure}

In Fig.~\ref{fig:OmgSpecMultiGam4Avg}, we plot 
gravitational wave power spectra from 
all simultaneous, exponential
and constant nucleation runs with $\gmStar\simeq 4$. 
We can see that they seem to be consistent, indicating that 
nucleation rate makes little difference to the power spectra as 
a function of $kR_*$.
We can therefore provide a 
fit for the gravitational wave power spectrum from 
collisions, applying to all nucleation histories.

The first two bins of the numerical power spectra contain very few modes and
are expected to be 
significantly affected by finite size effects. 
To produce our fit we shall use the largest simultaneous nucleation
simulation for $\gmStar\simeq 4$ with $\Nb=4096$ as this provides us with the largest
dynamic range. This is the only simulation in which we can resolve the
peak location after removing the
first two bins.
 
Even so, we do not have sufficient dynamic range to be
able to estimate the power law towards the IR. 
On causal grounds, though, 
it is expected that the IR power goes as $k^3$ \cite{Caprini:2009fx}. 
Our peak is somewhat broader
than previously seen in the envelope approximation. 

We find a fit of the following form
\begin{equation}\label{Eq:OurFit}
\frac{d\OmGW^\text{fit}}{d\text{ln}k}=\Omega^\text{fit}_\text{p}\frac{(a+b)^{c}\tilde{k}^b k^a}{(b \tilde{k}^{(a+b)/c}+ak^{(a+b)/c})^c}\text,
\end{equation}
where we fix $a=3$. Then we find that
\begin{align}
\Omega^\text{fit}_\text{p}&= (3.22\pm0.04) \times 10^{-3}\, (\Hc \Rstar \OmVac)^2\text, \\
\tilde{k}\Rstar&=3.20\pm0.04\text, \\
b &= 1.51\pm0.04,  \quad c = 2.18\pm0.15\text,
\end{align}
with errors taken from the covariance matrix of the fit. We plot our fit in Fig.~\ref{fig:OmgSpecMultiGam4Avg}.

We provide the fit in terms of the mean bubble separation $R_*$, which has a clear definition 
in all nucleation histories, 
and is related to the nucleation probability through equations 
(\ref{e:BubDenSim}), (\ref{e:BubDenExp}) and (\ref{e:BubDenCon}),
and the definition of $R_* = \nb^{-1/3}$.
For example, for exponential nucleation, 
\begin{equation}\label{eq:betaGivenRstar}
\beta=\frac{(8\pi)^{1/3}\vw}{\Rstar}\text.
\end{equation} 
Using Eq.~(\ref{eq:betaGivenRstar}) with $\vw=0.97$ we find that
\begin{equation}
\frac{\Om^\text{fit}_\text{p}}{\Om^\text{env}_\text{p}}=0.55\text,
\end{equation}
and
\begin{equation}
\frac{\tilde{k}_\text{fit}}{\tilde{k}_\text{env}}=1.0\text.
\end{equation}

We plot both our fit and also the fit from the envelope approximation
in Fig.~\ref{fig:OmgSpecMultiGam4Avg}.

\FloatBarrier
\section{Conclusions}\label{sec:Conclusions}

We have performed the largest scale lattice simulations of a pure
vacuum transition to date. In doing so we have been able to test the
envelope approximation's description of the resulting gravitational wave 
power spectrum, at high bubble wall Lorentz factors $\gmStar$ and for many bubbles. 
We have simulated three different bubble nucleation histories, 
where bubbles are either nucleated simultaneously,
with an exponentially increasing
nucleation rate, or with a constant rate.

In our simulations, the 
peak 
gravitational wave power 
has approximate agreement with the most recent envelope approximation fit \cite{Konstandin:2017sat},
to within a factor of two. 
The peak frequency  
in the envelope approximation fit 
has very good agreement with our results.

When the gravitational wave power is calculated using the envelope approximation's 
model of the actual bubbles of our simulation, 
the peak location is shifted 
towards slightly higher frequencies.

As we increase $\gmStar$ beyond $\gmStar\simeq 2$ we find that the
power law on the high frequency side of the peak 
becomes approximately $k^{-1.5}$, steeper than the $k^{-1}$ 
predicted by the envelope approximation.\footnote{
This steeper power law is closer to the $k^{-1.8}$ reported for 
two-bubble collisions \cite{Kosowsky:1991ua,Kosowsky:1992}, 
widely taken to be the envelope approximation 
prediction before the work of Ref.~\cite{Huber:2008}.
}
The power law on the low frequency side is consistent with 
the $k^3$ predicted by causality 
\cite{Caprini:2009fx,Durrer:2003ja},
but we do not have sufficient dynamic
range for an independent estimate. 
We provide a 3-parameter fit to our
results Eq.~(\ref{Eq:OurFit}).

In our simulations the overlap regions where bubbles
have recently collided have extended regions in which the scalar 
field has 
large amplitude oscillations around the true vacuum, 
even returning to the
false vacuum. 
These regions are not accounted for in the envelope approximation, 
and may be a source of its inaccuracy.
Large amplitude non-linear oscillations with wavelength 
of order the bubble wall width $l_0$ continue long 
after the bubbles finish colliding, which is also not included in the envelope approximation.
These oscillations 
source gravitational waves which 
lead to an additional bump in the UV of the power spectrum at a
frequency of order $l_0^{-1}$. 

In the early Universe,  the gravitational wave source 
will eventually diminish due to thermalisation and Hubble expansion. 
We find that even if
the bump continues to grow for as long as a Hubble time, $\Hc^{-1}$, 
the power spectrum from the oscillation phase will be subdominant to
that of bubble collisions providing that the mass of the scalar field is much less than the 
Planck mass.

In testing the envelope approximation and investigating the oscillatory phase of the scalar field, 
we have neglected the expansion of the Universe, and therefore 
the fit we provide strictly applies only to transitions in which the duration 
is much shorter than the Hubble time $H_*^{-1}$. There is more work 
to do to study the case where the Universe enters an inflationary phase 
before bubbles start nucleating.

\begin{acknowledgments}
We thank the Mainz Institute for Theoretical Physics (MITP) for
  its hospitality and support. We thank the LISA Cosmology Working Group for providing a forum to discuss our work, and extend further thanks to its coordinators Chiara Caprini and Germano Nardini.
  We are grateful to Stephan Huber, Ryusuke Jinno, and Kari Rummukainen for useful discussions, and 
  to Nicola Hopkins for important contributions to this project in its early stages. 
  Our simulations made use of the COSMOS
  Consortium supercomputer (within the DiRAC Facility jointly funded
  by STFC and the Large Facilities Capital Fund of BIS) and the
  Finnish Centre for Scientific Computing CSC. DC (ORCID ID 0000-0002-7395-7802) is
  supported by an STFC Studentship. MH (ORCID ID 0000-0002-9307-437X)
acknowledges support from the Science and Technology Facilities
Council (grant numbers ST/L000504/1 and ST/P000819/1). DJW (ORCID
  ID 0000-0001-6986-0517) was supported by Academy of Finland grant
  no.~286769 and the Research Funds of the University of Helsinki.
\end{acknowledgments}

\bibliography{scalar-only-biblio}

\end{document}